\documentclass[twocolumn,floatfix,aps,prd,superscriptaddress,nofootinbib]{revtex4-2}

\usepackage{iftex}
\ifPDFTeX
  \usepackage[utf8]{inputenc}
  \usepackage[T1]{fontenc}
  \usepackage{lmodern}
  \input{glyphtounicode}
\else
  \usepackage{fontspec}
\fi
\usepackage{graphicx}
\usepackage{amssymb}
\usepackage{amsmath}
\usepackage{xcolor}
\usepackage{hyperref}
\usepackage{textcase}
\usepackage{placeins}
\usepackage[protrusion=true,expansion=false]{microtype}

\AtBeginDocument{%
  \setcounter{dbltopnumber}{2}%
}

\makeatletter
\newcommand{\packfloatpages}{%
  \setlength{\@fptop}{0pt}%
  \setlength{\@fpsep}{12pt}%
  \setlength{\@fpbot}{0pt plus 1fil}%
  \setlength{\@dblfptop}{0pt}%
  \setlength{\@dblfpsep}{12pt}%
  \setlength{\@dblfpbot}{0pt plus 1fil}%
}
\makeatother

\graphicspath{{figures/}{../../../results/merger/figures/}}

\hypersetup{
    colorlinks=true,
    linkcolor=blue,
    filecolor=magenta,
    urlcolor=blue,
    citecolor=blue,
}
\newcommand{\clmDetRunsLoneThroat}{10}
\newcommand{\clmDetRunsSphericalKicks}{10}
\newcommand{\clmDetRunsQuadKicks}{6}
\newcommand{\clmDetRunsScalarReruns}{3}
\newcommand{\clmDetRunsVacuumBBH}{5}
\newcommand{\clmDetRunsTotal}{103}
\newcommand{\clmDetRunsBoostedThroat}{7}
\newcommand{\clmDetRunsSolveVerifications}{4}
\newcommand{\clmDetRunsMatchedRest}{9}
\newcommand{\clmDetRunsMatchedPlacement}{18}
\newcommand{\clmDetRunsHeadonModeThree}{8}
\newcommand{\clmDetRunsMergerChain}{7}
\newcommand{\clmDetRunsDTwelveArms}{2}
\newcommand{\clmDetRunsEgwScan}{8}
\newcommand{\clmDetKnobLoneLevelsLo}{2}
\newcommand{\clmDetKnobLoneLevelsHi}{4}
\newcommand{\clmDetKnobLoneL}{128}
\newcommand{\clmDetKnobLoneM}{0.5}
\newcommand{\clmDetKnobKickEpsLo}{10^{-3}}
\newcommand{\clmDetKnobKickEpsHi}{10^{-1}}
\newcommand{\clmDetKnobKickLevel}{5}
\newcommand{\clmDetKnobKickLevelLo}{3}
\newcommand{\clmDetKnobQuadLo}{0.005}
\newcommand{\clmDetKnobQuadHi}{0.05}
\newcommand{\clmDetKnobScalarLevel}{4}
\newcommand{\clmDetKnobScalarLLo}{64}
\newcommand{\clmDetKnobScalarLHi}{128}
\newcommand{\clmDetKnobScalarEps}{10^{-2}}
\newcommand{\clmDetKnobBoostPHi}{0.45}
\newcommand{\clmDetKnobMatchedALo}{1}
\newcommand{\clmDetKnobMatchedAHi}{3}
\newcommand{\clmDetKnobMatchedDLo}{12}
\newcommand{\clmDetKnobMatchedDHi}{18}
\newcommand{\clmDetKnobPlaceCsmDLo}{6}
\newcommand{\clmDetKnobPlaceCsmDHi}{48}
\newcommand{\clmDetKnobHoMotsLevelLo}{4}
\newcommand{\clmDetKnobHoMotsLevelHi}{6}
\newcommand{\clmDetKnobChainPTang}{0.10}
\newcommand{\clmDetKnobChainLevelLo}{3}
\newcommand{\clmDetKnobChainLevelHi}{7}
\newcommand{\clmDetKnobDTwelvePLo}{0.12}
\newcommand{\clmDetKnobDTwelvePHi}{0.25}
\newcommand{\clmDetKnobEgwPLo}{0.25}
\newcommand{\clmDetKnobEgwPHi}{0.90}
\newcommand{\clmDetGpuHours}{592}
\newcommand{\clmDetMemDevice}{80}
\newcommand{\clmDetAmrexVersion}{26.02}
\newcommand{\clmDetFFWindowLo}{0.893}
\newcommand{\clmDetFFWindowHi}{0.995}
\newcommand{\clmDetKerrFM}{0.0885}
\newcommand{\clmDetESpiralSpread}{1.1}
\newcommand{\clmDetEHeadon}{8.5\times10^{-3}}
\newcommand{\clmDetEHeadonSpread}{1.4}
\newcommand{\clmDetEThroatSpread}{3.6}
\newcommand{\clmDetETwinSpread}{10}
\newcommand{\clmDetSeedExampleC}{10^{-15}}
\newcommand{\clmDetOrbitPeriodLo}{75}
\newcommand{\clmDetOrbitPeriodHi}{93}
\newcommand{\clmDetLifePeriodsLo}{1.9}
\newcommand{\clmDetLifePeriodsHi}{2.4}
\newcommand{\clmDetDecadeUnitsLo}{12}
\newcommand{\clmDetCompanionLifeLo}{28}
\newcommand{\clmDetCompanionLifeHi}{29}
\newcommand{\clmNormPitShare}{99.99}
\newcommand{\clmDetRunsConvergence}{6}
\newcommand{\clmDetKnobConvArmLevel}{4}

\newcommand{\clmHeadonSeparation}{8}
\newcommand{\clmSpiralMomentumProd}{0.12}
\newcommand{\clmSpiralMomentumTwentyFive}{0.25}
\newcommand{\clmSpiralMomentumFlyby}{0.45}
\newcommand{\clmSpiralBoundaryLow}{90}
\newcommand{\clmSpiralBoundaryHigh}{120}
\newcommand{\clmControlChiDigits}{four}
\newcommand{\clmControlByteRows}{633}
\newcommand{\clmHeadonGwShare}{4.7}
\newcommand{\clmFlybyTrustRetarded}{67.6}
\newcommand{\clmFarMassIso}{-4.8105}
\newcommand{\clmFarChargeIso}{3.0344}
\newcommand{\clmMatchedRminDev}{0.06}
\newcommand{\clmVolumeIdentityExact}{0.14}
\newcommand{\clmEnergyLawDevNear}{3}
\newcommand{\clmEnergyLimitAttract}{+4}
\newcommand{\clmEnergyLimitRepel}{-6}
\newcommand{\clmMatchedOneBodyTol}{6\times10^{-6}}
\newcommand{\clmPlaceTabRiso}{3.890}
\newcommand{\clmPlaceTabMiso}{-4.810}
\newcommand{\clmPlaceTabAiso}{1.001}
\newcommand{\clmPlaceTabRmat}{3.892}
\newcommand{\clmPlaceTabMmat}{-4.810}
\newcommand{\clmPlaceTabAmat}{2.362}
\newcommand{\clmEnergyLambdaAttractEight}{0.857}
\newcommand{\clmEnergyBindAttractEight}{+0.357}
\newcommand{\clmEnergyLawAttractEight}{+0.367}
\newcommand{\clmEnergyLambdaRepelEight}{0.906}
\newcommand{\clmEnergyBindRepelEight}{-0.619}
\newcommand{\clmEnergyLawRepelEight}{-0.616}
\newcommand{\clmEnergyLambdaAttractTwelve}{0.906}
\newcommand{\clmEnergyBindAttractTwelve}{+0.269}
\newcommand{\clmEnergyLawAttractTwelve}{+0.274}
\newcommand{\clmEnergyLambdaRepelTwelve}{0.931}
\newcommand{\clmEnergyBindRepelTwelve}{-0.437}
\newcommand{\clmEnergyLawRepelTwelve}{-0.433}
\newcommand{\clmEnergyLambdaAttractSixteen}{0.931}
\newcommand{\clmEnergyBindAttractSixteen}{+0.214}
\newcommand{\clmEnergyLawAttractSixteen}{+0.217}
\newcommand{\clmEnergyLambdaRepelSixteen}{0.946}
\newcommand{\clmEnergyBindRepelSixteen}{-0.338}
\newcommand{\clmEnergyLawRepelSixteen}{-0.335}
\newcommand{\clmEnergyLambdaAttractTwentyFour}{0.955}
\newcommand{\clmEnergyBindAttractTwentyFour}{+0.151}
\newcommand{\clmEnergyLawAttractTwentyFour}{+0.152}
\newcommand{\clmEnergyLambdaRepelTwentyFour}{0.962}
\newcommand{\clmEnergyBindRepelTwentyFour}{-0.233}
\newcommand{\clmEnergyLawRepelTwentyFour}{-0.231}
\newcommand{\clmEnergyLambdaAttractFortyEight}{0.979}
\newcommand{\clmEnergyBindAttractFortyEight}{+0.080}
\newcommand{\clmEnergyLawAttractFortyEight}{+0.080}
\newcommand{\clmEnergyLambdaRepelFortyEight}{0.980}
\newcommand{\clmEnergyBindRepelFortyEight}{-0.121}
\newcommand{\clmEnergyLawRepelFortyEight}{-0.120}
\newcommand{\clmHeadonCsmAdmMass}{2.3573}
\newcommand{\clmHeadonCsmSeamOne}{35}
\newcommand{\clmHeadonCsmSeamTwo}{50.8}
\newcommand{\clmHeadonCsmNanTime}{38.845}
\newcommand{\clmHeadonCsmMotsTime}{18}
\newcommand{\clmHeadonCsmMotsRadius}{5.634}
\newcommand{\clmHeadonCsmMotsMass}{2.817}
\newcommand{\clmHeadonCsmEndRadius}{4.7787}
\newcommand{\clmHeadonCsmEndOverSchw}{1.4}
\newcommand{\clmHeadonCsmEndMass}{2.3894}
\newcommand{\clmHeadonCsmPenrose}{19.5}
\newcommand{\clmHeadonCsmBornArea}{1.05}
\newcommand{\clmHeadonCsmScalarImplied}{0.052}
\newcommand{\clmHeadonCsmBirthSep}{3.5}
\newcommand{\clmHeadonScanLate}{4}
\newcommand{\clmHeadonScanSmall}{11}
\newcommand{\clmHeadonCsmShrinkArea}{28}
\newcommand{\clmHeadonCsmRegrow}{0.34}
\newcommand{\clmHeadonFirstLawMeasured}{-4.1\times10^{-3}}
\newcommand{\clmHeadonFirstLawPhantom}{-7.0\times10^{-3}}
\newcommand{\clmHeadonFirstLawShear}{-2.5\times10^{-3}}
\newcommand{\clmHeadonCsmFluxFold}{69}
\newcommand{\clmHeadonCsmSlope}{7.7\times10^{-4}}
\newcommand{\clmMergerRemnantAgree}{2.2}
\newcommand{\clmFlybyLbfClosest}{2.32}
\newcommand{\clmPFortyFiveClosest}{2.18}
\newcommand{\clmPFortyFiveCircFrac}{90}
\newcommand{\clmPSixtyCircFrac}{120}
\newcommand{\clmFlybyLbfTrust}{63.3}
\newcommand{\clmPFortyFiveTrust}{67.6}
\newcommand{\clmDTwelveInwardP}{0.25}
\newcommand{\clmDTwelveInflTwentyFour}{72}
\newcommand{\clmDTwelveStallSep}{2.2}
\newcommand{\clmPTwelveContactTime}{37}
\newcommand{\clmPTwelveMotsFrom}{51}
\newcommand{\clmPTwelveTrust}{56.5}
\newcommand{\clmRaceKickTwelve}{0.4}
\newcommand{\clmRaceKickEight}{1.2}
\newcommand{\clmRaceFoldsHeadon}{3.5}
\newcommand{\clmRaceFoldsSix}{2.5}
\newcommand{\clmRaceFoldsTwelve}{9}
\newcommand{\clmDSixP}{0.10}
\newcommand{\clmDSixCircFrac}{14}
\newcommand{\clmDampOnTime}{36}
\newcommand{\clmDampOffReach}{34}
\newcommand{\clmMergerMotsTime}{13}
\newcommand{\clmMergerMotsRadius}{5.60}
\newcommand{\clmMergerShrinkR}{12.8}
\newcommand{\clmMergerEndRadius}{4.883}
\newcommand{\clmMergerBirthSep}{3.6}
\newcommand{\clmMergerLegOneStop}{26.26}
\newcommand{\clmMergerLvlSixDeath}{29.4}
\newcommand{\clmMergerLvlSevenDeath}{27.5}
\newcommand{\clmMergerSigDeath}{30.5}
\newcommand{\clmMergerChiStop}{36.7}
\newcommand{\clmMergerSeamOne}{25}
\newcommand{\clmMergerSeamTwo}{30}
\newcommand{\clmMergerSeamThree}{35}
\newcommand{\clmMergerWallLeadLo}{15}
\newcommand{\clmMergerWallLeadHi}{18}
\newcommand{\clmPSixtyExtDeath}{55.5}
\newcommand{\clmPNinetyDeath}{45.3}
\newcommand{\clmPNinetyKLo}{93}
\newcommand{\clmPNinetyKHi}{172}
\newcommand{\clmPNinetyLvlFiveDeath}{45.28}
\newcommand{\clmEgwPSixty}{0.60}
\newcommand{\clmEgwPNinety}{0.90}
\newcommand{\clmHeadonCsmWallLead}{21}
\newcommand{\clmPSixtyTrust}{80}
\newcommand{\clmBoostFaceBias}{2.8}
\newcommand{\clmDampOffPsi}{2\times10^{-7}}

\newcommand{\clmThroatA}{2}
\newcommand{\clmThroatM}{1}
\newcommand{\clmRstar}{3.8895}
\newcommand{\clmThroatCoordRadius}{1.618}
\newcommand{\clmChiThroatLo}{0.15}
\newcommand{\clmChiThroatHi}{0.25}
\newcommand{\clmScalarQ}{5}
\newcommand{\clmScalarQRatio}{3/2}
\newcommand{\clmSeedArealFactor}{2}
\newcommand{\clmMergerSign}{-1}
\newcommand{\clmPairSeparation}{12}
\newcommand{\clmBoxL}{64}
\newcommand{\clmBaseN}{128}
\newcommand{\clmTimeStepFactor}{0.02}
\newcommand{\clmKOsigma}{0.1}
\newcommand{\clmShiftEta}{1}
\newcommand{\clmRegridSteps}{16}
\newcommand{\clmCczKappaOne}{3}
\newcommand{\clmCczKappaThree}{1}
\newcommand{\clmDxBase}{0.5}
\newcommand{\clmFinestHalfWidth}{4}
\newcommand{\clmProdLevels}{three}
\newcommand{\clmFinestDx}{1/16}
\newcommand{\clmResArmsLevelHi}{7}
\newcommand{\clmBigBoxL}{128}
\newcommand{\clmSpongeStrength}{4}
\newcommand{\clmSpongePower}{fourth}
\newcommand{\clmChiFloor}{10^{-8}}
\newcommand{\clmLapseFloor}{10^{-10}}
\newcommand{\clmChiFloorVariant}{10^{-20}}
\newcommand{\clmChiRhsFloor}{10^{-8}}
\newcommand{\clmExtractRadiusTen}{10}
\newcommand{\clmExtractRadiusEighteen}{18}
\newcommand{\clmExtractRadiusTwenty}{20}
\newcommand{\clmExtractRadiusTwentyEight}{28}
\newcommand{\clmExtractRadiusFortyFour}{44}
\newcommand{\clmShapeSystLone}{3}
\newcommand{\clmFlowFinderLmax}{6}
\newcommand{\clmFlowFinderLmaxAlt}{8}
\newcommand{\clmFlowFinderSchw}{0.24}
\newcommand{\clmPsiPipelines}{0.4}
\newcommand{\clmHoldTimeLevelThree}{35}
\newcommand{\clmHoldTolLevelThree}{0.1}
\newcommand{\clmHoldTimeOnePct}{44}
\newcommand{\clmHoldTolOnePct}{1}
\newcommand{\clmLevelFourDigits}{four}
\newcommand{\clmOnsetShiftLo}{5.6}
\newcommand{\clmOnsetShiftHi}{8.7}
\newcommand{\clmLevelTwoDeath}{24.2}
\newcommand{\clmHalfMassM}{0.5}
\newcommand{\clmHalfMassHoldTol}{0.1}
\newcommand{\clmHalfMassHoldTime}{17}
\newcommand{\clmHalfMassDeath}{26.3}
\newcommand{\clmTauLevelThree}{5.12}
\newcommand{\clmTauLevelFour}{5.23}
\newcommand{\clmTauFitDev}{4}
\newcommand{\clmTauRounded}{5.2}
\newcommand{\clmTauLevelFourRounded}{5.1}
\newcommand{\clmProperEfoldThree}{0.76}
\newcommand{\clmProperEfoldFour}{0.77}
\newcommand{\clmThroatLapse}{0.575}
\newcommand{\clmGgsAgreement}{0.2}
\newcommand{\clmCrossingMean}{13.0}
\newcommand{\clmKickThreeLead}{15}
\newcommand{\clmDressedLeadLevelThree}{14}
\newcommand{\clmMatchedDelayLo}{11}
\newcommand{\clmMatchedDelayHi}{12}
\newcommand{\clmHorizonTimeLevelFour}{11}
\newcommand{\clmCrossingLevelFour}{12.99}
\newcommand{\clmCrossingLevelThree}{13.01}
\newcommand{\clmHorizonTimeEpsTwo}{11}
\newcommand{\clmHorizonTimeEpsThree}{25}
\newcommand{\clmLawEpsThree}{22.8}
\newcommand{\clmHorizonRadiusEpsTwo}{3.88}
\newcommand{\clmHorizonRadiusEpsThree}{3.81}
\newcommand{\clmHorizonTimeNoise}{61}
\newcommand{\clmNoiseSeed}{6\times10^{-7}}
\newcommand{\clmQuietLifetime}{19}
\newcommand{\clmPhysRstarThirty}{172}
\newcommand{\clmPhysRstarThousand}{5.7\times10^{3}}
\newcommand{\clmPhysRstarMillion}{5.7\times10^{6}}
\newcommand{\clmPhysRstarBillion}{38}
\newcommand{\clmPhysTauThirty}{0.76}
\newcommand{\clmPhysTauThousand}{25}
\newcommand{\clmPhysTauMillion}{25}
\newcommand{\clmPhysTauBillion}{7.0}
\newcommand{\clmPhysEpsTwoThirty}{1.6}
\newcommand{\clmPhysEpsTwoThousand}{54}
\newcommand{\clmPhysEpsTwoMillion}{54}
\newcommand{\clmPhysEpsTwoBillion}{15}
\newcommand{\clmPhysEpsThreeThirty}{3.7}
\newcommand{\clmPhysEpsThreeThousand}{0.12}
\newcommand{\clmPhysEpsThreeMillion}{2.1}
\newcommand{\clmPhysEpsThreeBillion}{1.4}
\newcommand{\clmPhysNoiseThirty}{9.0}
\newcommand{\clmPhysNoiseThousand}{0.30}
\newcommand{\clmPhysNoiseMillion}{5.0}
\newcommand{\clmPhysNoiseBillion}{3.5}
\newcommand{\clmRstarShort}{3.89}
\newcommand{\clmUnkickedHorizonRStart}{3.23}
\newcommand{\clmUnkickedHorizonREnd}{2.37}
\newcommand{\clmUnkickedHorizonSpan}{39}
\newcommand{\clmKickedRunsClean}{100}
\newcommand{\clmPureQuadAmp}{10^{-2}}
\newcommand{\clmPureQuadAmpCaption}{0.01}
\newcommand{\clmPureQuadMotsTime}{33}
\newcommand{\clmPureQuadMotsRStart}{3.80}
\newcommand{\clmPureQuadMotsREnd}{2.50}
\newcommand{\clmPureQuadTrust}{60}
\newcommand{\clmGwSingleAgreePct}{3}
\newcommand{\clmPureQuadDeparture}{42}
\newcommand{\clmCollapseShells}{128}
\newcommand{\clmRegrowthFloorTime}{47}
\newcommand{\clmFourBoxL}{512}
\newcommand{\clmFourLevel}{5}
\newcommand{\clmFourWall}{218}
\newcommand{\clmFourRStart}{3.81}
\newcommand{\clmFourRRatio}{3.8}
\newcommand{\clmFourSHRate}{1.22}
\newcommand{\clmFourSHWinHi}{40}
\newcommand{\clmFourSHLocalLo}{1.0}
\newcommand{\clmFourSHLocalHi}{1.3}
\newcommand{\clmFourTauLate}{5}
\newcommand{\clmFourLapseStart}{0.57}
\newcommand{\clmFourLapseEnd}{0.015}
\newcommand{\clmSHMassless}{1.1}
\newcommand{\clmFourLinearRate}{1.30}
\newcommand{\clmFourHorizonR}{73}
\newcommand{\clmFourHorizonTime}{212}
\newcommand{\clmSignSettleWindow}{3.5}
\newcommand{\clmSignWindowEnd}{10.5}
\newcommand{\clmSignWindowClosed}{4.6}
\newcommand{\clmMatchedSignRatio}{1.462}
\newcommand{\clmMatchedSignRatioErr}{0.022}
\newcommand{\clmMatchedOffsetDelta}{2.65}
\newcommand{\clmCountLikeSignedPairs}{Four}
\newcommand{\clmDLadderEighteen}{0.2406}
\newcommand{\clmFdSquaredLo}{69.0}
\newcommand{\clmFdSquaredHi}{78.0}
\newcommand{\clmOffsetPrediction}{0.242}
\newcommand{\clmInverseSquarePrediction}{0.213}
\newcommand{\clmWidthExponentLo}{1.2}
\newcommand{\clmWidthExponentHi}{1.4}
\newcommand{\clmThroatGammaOne}{0.5}
\newcommand{\clmGgsMatched}{0.758}
\newcommand{\clmLinT}{0.7584}
\newcommand{\clmLinTau}{5.13}
\newcommand{\clmGgsAgreementFour}{1.9}
\newcommand{\clmThroatOwnMass}{1.94}
\newcommand{\clmSeedDefectPeak}{0.93}
\newcommand{\clmSeedDefectPeakHi}{0.94}
\newcommand{\clmSeedBagThreshold}{0.025}
\newcommand{\clmSeedDefectBox}{0.090}
\newcommand{\clmCeilingPlusMotsTime}{1}
\newcommand{\clmCeilingMinusMotsTime}{6}
\newcommand{\clmRegrowthShrinkPct}{40}
\newcommand{\clmInfStoreDeficit}{0.94}
\newcommand{\clmInfShedQuiet}{0.03}
\newcommand{\clmFourFinestDx}{1/16}
\newcommand{\clmFourLinearGap}{6}
\newcommand{\clmBoostAxisResidual}{6\times10^{-5}}
\newcommand{\clmBoostSolveW}{4.5\times10^{-6}}
\newcommand{\clmBoostRestRepro}{10^{-5}}
\newcommand{\clmBoostPairFar}{10^{-5}}
\newcommand{\clmBoostSpeed}{0.41}
\newcommand{\clmBoostMovingOnePct}{36.8}
\newcommand{\clmBoostMovingTauLo}{5.0}
\newcommand{\clmBoostMovingTauHi}{6.1}
\newcommand{\clmBoostGamma}{1.10}
\newcommand{\clmBoostDilatedTau}{5.6}
\newcommand{\clmBoostMovingDeath}{44.67}
\newcommand{\clmFreezeRadiusFull}{0.3}
\newcommand{\clmFreezeRadiusStart}{0.8}
\newcommand{\clmDampLapseStart}{3\times10^{-2}}
\newcommand{\clmScatterSepStart}{11.9}
\newcommand{\clmScatterSepSixty}{26.4}
\newcommand{\clmScatterRminStart}{3.88}
\newcommand{\clmScatterRminThirty}{4.78}
\newcommand{\clmScatterMotsLast}{16}

\newcommand{\clmGwPeakSpiral}{2.1\times10^{-2}}
\newcommand{\clmGwPeakThroat}{2.9\times10^{-3}}
\newcommand{\clmGwEnergyFlyby}{7.6\times10^{-2}}
\newcommand{\clmGwEnergySpiral}{4.2\times10^{-3}}
\newcommand{\clmGwEnergyTwin}{2.4\times10^{-3}}
\newcommand{\clmGwEnergyThroat}{2.5\times10^{-5}}
\newcommand{\clmGwSpreadFlyby}{1.9}
\newcommand{\clmGwVacPassRatioLo}{73}
\newcommand{\clmGwVacPassRatioHi}{74}
\newcommand{\clmGwSpeedTwinPeak}{0.57}
\newcommand{\clmGwSpeedTwin}{0.82}
\newcommand{\clmGwLigoMass}{30}
\newcommand{\clmGwLigoDist}{10}
\newcommand{\clmGwIscoHz}{147}
\newcommand{\clmGwKerrHz}{599}
\newcommand{\clmGwTwinPeakFM}{0.065}
\newcommand{\clmGwLitHeadonInfall}{5.5\times10^{-4}}
\newcommand{\clmGwLitCircular}{4.84\times10^{-2}}
\newcommand{\clmGwSeedStrong}{0.05}
\newcommand{\clmGwThroatOverControl}{44}
\newcommand{\clmGwThroatSpeedLo}{0.8}
\newcommand{\clmGwThroatSpeedHi}{1.0}
\newcommand{\clmGwLinDevLo}{1}
\newcommand{\clmGwLinDevHi}{16}
\newcommand{\clmGwSeedWeak}{0.005}
\newcommand{\clmGwPureVsKicked}{2}
\newcommand{\clmGwThroatPeriod}{20.6}
\newcommand{\clmGwThroatMassFloor}{1.16}
\newcommand{\clmGwThroatMassEnd}{1.21}
\newcommand{\clmGwThroatMassStart}{1.47}
\newcommand{\clmGwHeadonRadiusA}{10}
\newcommand{\clmGwHeadonSpeed}{0.92}
\newcommand{\clmGwTwinFreqLo}{299}
\newcommand{\clmGwTwinFreqHi}{473}
\newcommand{\clmGwScalarRadiusOuter}{30}
\newcommand{\clmGwScalarFluxFactor}{0.94}
\newcommand{\clmGwScalarFluxCorr}{6}
\newcommand{\clmGwScalarDipoleShare}{6.2}
\newcommand{\clmGwScalarRadiusInner}{14}
\newcommand{\clmGwScalarCut}{60}
\newcommand{\clmGwScalarEnergyFlyby}{4.7\times10^{-2}}
\newcommand{\clmGwHeadonRadiusB}{14}
\newcommand{\clmGwHeadonRadiusC}{18}
\newcommand{\clmGwKickedLevel}{4}
\newcommand{\clmGwKickedEps}{+10^{-2}}
\newcommand{\clmGwKickedMotsTime}{11}
\newcommand{\clmGwKickedEfoldLo}{6}
\newcommand{\clmGwKickedEfoldHi}{7}
\newcommand{\clmGwSingleLevel}{4}
\newcommand{\clmGwSingleEfold}{4.6}
\newcommand{\clmGwSingleBox}{128}
\newcommand{\clmGwSingleCountTime}{80}
\newcommand{\clmGwEnvWindow}{25}
\newcommand{\clmGwFluxSignStretch}{13}
\newcommand{\clmGwDrawFloorFrac}{14}
\newcommand{\clmGwDrawFlybyRetarded}{67.6}
\newcommand{\clmGwThroatRingMass}{1.226}
\newcommand{\clmGwThroatRingMassErr}{0.009}
\newcommand{\clmGwLitScatter}{2\times10^{-2}}
\newcommand{\clmGwThroatEnergyCut}{58}
\newcommand{\clmGwHeadonNoiseGate}{80}
\newcommand{\clmEgwEnergyTwoFive}{3.53\times10^{-2}}
\newcommand{\clmEgwEnergyFortyFive}{7.6\times10^{-2}}
\newcommand{\clmEgwEnergySixty}{4.1\times10^{-2}}
\newcommand{\clmEgwGateSixty}{12}
\newcommand{\clmEgwBoundaryOverScatter}{2.2}
\newcommand{\clmEgwBoundaryOverPlunge}{1.9}
\newcommand{\clmHeadonCsmGwEnergy}{8.5\times10^{-3}}
\newcommand{\clmBbhHeadonGwRatio}{14}
\newcommand{\clmBbhHeadonPeakLag}{28}
\newcommand{\clmGwSpeedLow}{0.82}
\newcommand{\clmGwSpeedHigh}{1.00}
\newcommand{\clmGwHeadonHalfLo}{15.5}
\newcommand{\clmGwHeadonHalfHi}{18.8}
\newcommand{\clmGwScalarRatioFull}{0.56}
\newcommand{\clmJunkShareMax}{2.7}
\newcommand{\clmScalarNearSlope}{2.5}
\newcommand{\clmScalarPrearchShare}{14}
\newcommand{\clmScalarArchRatio}{0.48}
\newcommand{\clmScalarQnmFreqDev}{1}
\newcommand{\clmScalarQnmDampDev}{10}
\newcommand{\clmMemoryRatioLo}{0.34}
\newcommand{\clmMemoryRatioHi}{0.46}
\newcommand{\clmHeadonQnmDev}{3}
\newcommand{\clmGwScalarGateOuter}{97.6}
\newcommand{\clmGwCensMergerTau}{22}
\newcommand{\clmGwVacMergerPeakRatio}{3.0}
\newcommand{\clmHeadonAxisRatioDev}{2}
\newcommand{\clmEgwTrustTwoFive}{63.3}
\newcommand{\clmConvOrderDataLo}{2.2}
\newcommand{\clmConvOrderDataHi}{2.3}
\newcommand{\clmConvOrderEvolLo}{3.0}
\newcommand{\clmConvOrderEvolHi}{3.3}
\newcommand{\clmConvPsiLevelDiff}{0.4}
\newcommand{\clmConvPsiWaveZoneDiff}{0.15}
\newcommand{\clmNearZoneHeadonInner}{28}
\newcommand{\clmNearZoneHeadonOuterSpread}{1.5}
\newcommand{\clmNearZoneAmpExcess}{13}
\newcommand{\clmConvFzDxLo}{16}
\newcommand{\clmConvFzDxMid}{32}
\newcommand{\clmConvFzDxHi}{64}
\newcommand{\clmConvFzPsiDiff}{0.55}
\newcommand{\clmConvFzPsiDiffFine}{0.54}
\newcommand{\clmConvFzMotsTime}{18}
\newcommand{\clmConvFzOrder}{1.7}
\newcommand{\clmConvFzHamOrder}{1.7}
\newcommand{\clmMassTzeroExcessLo}{2.4}
\newcommand{\clmMassTzeroExcessHi}{3.2}
\newcommand{\clmFarzoneBoxL}{512}
\newcommand{\clmFarzoneOuterR}{150}

\newcommand{\tcell}[2]{#2}
\begin{document}

\title{\texorpdfstring{Binary Ellis--Bronnikov wormholes in 3+1 numerical relativity: mergers, fly-bys and gravitational waves}{Binary Ellis-Bronnikov wormholes in 3+1 numerical relativity: mergers, fly-bys and gravitational waves}}

\author{Nikita M. Shirokov}
\email{shirokov.nm@phystech.edu}
\affiliation{\textit{Moscow Institute of Physics and Technology, Moscow, Russia}}
\author{Ilya Nachevsky}
\email{ommnammashivaya@gmail.com}
\affiliation{\textit{Higher School of Economics, Moscow, Russia}}
\date{\today}

\begin{abstract}
We present what are, to our knowledge, the first three-dimensional numerical-relativity evolutions of wormhole binaries: pairs of massive Ellis--Bronnikov wormholes held open by a phantom scalar field, whose energy density is nowhere positive. We construct constraint-satisfying initial data for single throats and for pairs, at rest and with orbital momentum, and evolve them with adaptive mesh refinement. An isolated throat is unstable: depending on the sign of a small perturbation, it collapses to a black hole or inflates, at the rate that linear perturbation theory predicts to within $\clmGgsAgreementFour\%$. Two throats at rest repel when their scalar charges have the same sign and attract when the signs are opposite, with a strength ratio close to the scalar force law's prediction, so only opposite-signed pairs merge. A head-on collision forms a common horizon around both throats while they are still open. The horizon then loses $\clmHeadonCsmShrinkArea\%$ of its area as it absorbs the scalar field, and approaches the Schwarzschild radius of the pair's mass. Pairs that fly past each other form no horizon, and both throats inflate. Every encounter emits a short burst of gravitational waves and a scalar dipole wave of comparable, negative energy. An isolated collapsing throat rings at the quasinormal frequency of the black hole it becomes. No throat survives more than a few orbital periods, even when its initial perturbation is tiny, so these binaries do not inspiral. The results hold for one phantom-scalar model, equal masses, a single throat compactness and no spin.
\end{abstract}

\maketitle

\section{Introduction}
\label{sec:intro}

Wheeler's spacetime foam is a sea of handles---wormholes that open and close on the Planck scale~\cite{wheeler1955,wheeler1957,misner_wheeler1957}. Several mechanisms promote them to macroscopic size: inflation stretches a Planck-scale Lorentzian wormhole with the background~\cite{roman1993}; bubble and domain-wall cosmologies produce post-inflationary regions joined through wormholes~\cite{garriga2016,deng2017,deng_vilenkin2017,gouttenoire2024}; and a gas of relic wormholes has been studied as a cosmology in its own right~\cite{kirillov2008,kirillov2011}. The bubble and wall wormholes are not traversable and pinch off into black holes~\cite{deng2017,deng_vilenkin2017}. None of this work evolves a traversable relic. This paper asks what one does, and what two do to each other.

Two classical facts govern the dynamics. First, the simplest evolvable traversable wormhole is unstable~\cite{gonzalez2009lin,bronnikov2011,cremona2019}. It is the Ellis--Bronnikov wormhole held open by a phantom scalar~\cite{ellis1973,bronnikov1973,morris_thorne1988,visser1995}, which Ellis called a drainhole; we study its massive form, with nonzero Arnowitt--Deser--Misner (ADM) mass. Depending on the sign of the perturbation, the throat collapses to a black hole or inflates, on a timescale set by the throat radius~\cite{hayward1999,shinkai2002,gonzalez2009,doroshkevich2009,hayward2009,novikov2019}. Evolutions in 3+1 dimensions confirm this nonlinearly for the massless throat~\cite{shirokov2026}. As a quantum field coupled to gravity a ghost destabilises the vacuum unless its effective theory is cut off at low energy with Lorentz violation~\cite{cline2004}; we treat it as a classical phenomenological model. Second, Geroch's theorem forbids classical topology change~\cite{geroch1967,tipler1977}: two wormholes cannot fuse into one, and none can be created classically. Creation is quantum~\cite{hawking1988}, so our evolutions start at $t=0$ from the newly created state.

\begin{figure*}[t]
\centering
\includegraphics[width=0.88\textwidth]{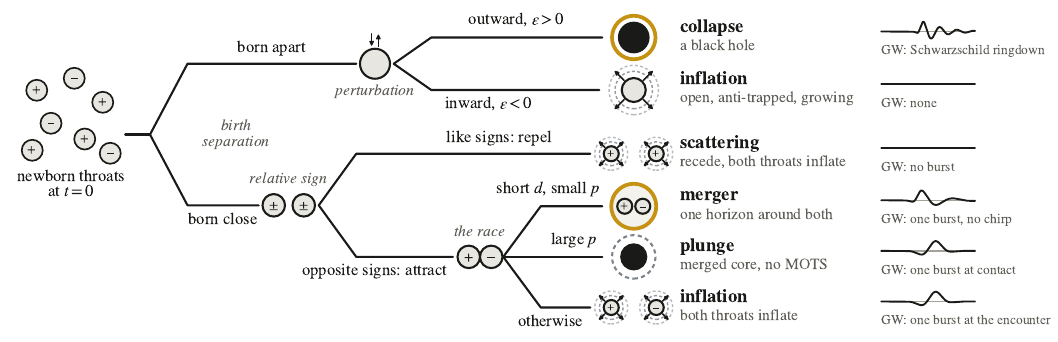}
\caption{Schematic of the end states of newly formed throats, whatever their origin. An isolated throat's perturbation sign selects collapse or inflation (Sec.~\ref{sec:single:branching}). A pair's relative charge sign selects repulsion or attraction (Sec.~\ref{sec:pair:sign}). The throats of a repelling pair inflate as it recedes (Sec.~\ref{sec:pair:recede}). An attracting pair merges if its common horizon forms early in the throats' inflation (Sec.~\ref{sec:spiral:race}); in a fast plunge it collapses to a merged core around which no common horizon is found (Sec.~\ref{sec:spiral:capture}). Gold: horizons; right: sketched wave signals. Two end states give a black hole after a single burst without a chirp, the fast plunge gives a merged core, and in every other end state the throats inflate.}
\label{fig:fates}
\end{figure*}

These facts constrain how wormhole pairs can arise. No astrophysical assembly channel exists, because a throat decays faster than dynamical friction or radiation reaction could pair it. A primordial pair born from foam needs no assembly: it is born at a separation comparable to its throats, so its birth and its collision coincide. Its evolution is then a race between collision and throat decay. The scalar field sets a charge rule: a phantom field repels like charges and attracts opposite ones, so like-signed pairs scatter and opposite-signed pairs fall together. The instability sets a second rule: the sign of an isolated throat's perturbation selects inflation or collapse to a black hole, a channel to primordial black holes beside the standard ones~\cite{zeldovich1967,carr_hawking1974,carr2021}. We map the resulting end states (merger, scattering, collapse, inflation, and a fast plunge to a merged core that the grid does not resolve) across separation, perturbation and relative sign (Fig.~\ref{fig:fates}). Our binary has three asymptotic ends, since each of its two wormholes leads to its own far universe. A foam handle is one wormhole whose two mouths carry opposite charge, so our opposite-signed pairs share its charges but not its topology. A merger does not fuse two wormholes into one: both throats end behind one common horizon.

The observable is not a chirp but a population of bursts from the merger epoch, redshifted into whatever band the mass scale selects~\cite{caprini2018,nanograv2023,lisa2017,lvk2021}. No throat survives more than milliseconds to days (Sec.~\ref{sec:single:lifetime}), so neither LIGO nor LISA~\cite{aligo2010,robson2019,thrane2013} can see a surviving wormhole. A population born in the early universe collapses or inflates soon after birth, and its bursts are redshifted far below both bands. A drainhole signal in these bands would need throats born that late, which no known mechanism supplies. The epoch-independent result is the energy radiated per event and its spectrum, in units of the source's ADM mass (Sec.~\ref{sec:ligo:burst}). The waveforms are extracted at finite radius in the near zone, without extrapolation to null infinity, and each energy carries the spread across extraction spheres as its error (Sec.~\ref{sec:gw}).

Section~\ref{sec:model} presents the model and the initial data. Section~\ref{sec:setup} describes the numerical methods. Sections~\ref{sec:single}--\ref{sec:spiral} follow, in turn, the single wormhole, two throats at rest, the head-on merger and the orbital encounters. Section~\ref{sec:gw} treats the radiation of every encounter. Section~\ref{sec:ligo} discusses observational signatures. Section~\ref{sec:discussion} draws the consequences. Section~\ref{sec:repro} covers reproducibility and cost. Appendix~\ref{app:health} gives the convergence tests, and Appendix~\ref{app:supp} supplementary measurements. All evolutions use \texttt{GRTeclyn}~\cite{grteclyn}, the AMReX-based~\cite{amrex} GPU port of \texttt{GRChombo}~\cite{grchombo}, continuing Refs.~\cite{shirokov2026,shirokov2026b}. We use $G=c=1$ and measure masses, lengths and times in units of the drainhole mass parameter $m$ (Sec.~\ref{sec:model:drainhole}). The results hold for a single minimally coupled phantom scalar, equal masses, one throat compactness ($m/a=1/2$) and no spin.

\section{The model}
\label{sec:model}

\subsection{Field equations}
\label{sec:model:field}

The matter is a single real, massless phantom (ghost) scalar: gravity couples to minus the canonical stress tensor,
\begin{equation}
T_{ab} = -\Bigl(\nabla_a\phi\,\nabla_b\phi - \tfrac12 g_{ab}\,\nabla_c\phi\nabla^c\phi\Bigr),
\qquad V(\phi)\equiv 0,
\label{eq:phantom}
\end{equation}
so the energy density seen by normal observers, $\rho=-\tfrac12\bigl(\Pi^2+\gamma^{ij}\partial_i\phi\,\partial_j\phi\bigr)$ with $\Pi=(\partial_t\phi-\beta^i\partial_i\phi)/\alpha$, where $\alpha$ is the lapse, $\beta^i$ the shift and $\gamma_{ij}$ the spatial metric, is non-positive everywhere. The null energy condition fails wherever $\phi$ varies, and this exotic stress holds a throat open. The field itself obeys the canonical wave equation $\Box\phi=0$, evolved in first-order form in $\phi$ and $\Pi$. Table~\ref{tab:notation} collects the notation.

\begin{table}[t]
\caption{Notation. We use $G=c=1$ and units of the drainhole mass parameter $m$.}
\label{tab:notation}
\scriptsize
\begin{ruledtabular}
\begin{tabular}{lp{0.6\columnwidth}}
symbol & meaning \\
\hline
$\alpha$, $\beta^i$, $\gamma_{ij}$ & lapse, shift, spatial metric \\
$K_{ij}$, $K$ & extrinsic curvature and its trace \\
$\chi$, $h_{ij}$ & conformal factor and conformal metric, $\gamma_{ij}=h_{ij}/\chi$ \\
$\Psi$, $\psi$ & conformal factors of the initial data, $\gamma_{ij}=\Psi^4\delta_{ij}$, $\Psi=e^{-u/2}\psi$ \\
$\phi$, $\Pi$ & phantom scalar field and its momentum \\
$\mathcal H$, $\mathcal M_i$ & Hamiltonian and momentum constraints \\
$a$, $m$ & throat scale and mass parameter of the drainhole \\
$r$, $r_t$ & isotropic radius about a throat centre; its value at the throat \\
$R$, $R_\star$ & areal radius; throat radius of the static drainhole \\
$q$, $Q$ & scalar charge; scalar-to-gravitational force ratio, Eq.~\eqref{eq:force} \\
$d$, $\sigma$ & coordinate separation; relative sign of the two scalar charges \\
$p$, $v$, $\gamma$ & momentum per throat, boost speed, Lorentz factor \\
$M_{\rm ADM}$, $M_{\rm far}$ & ADM mass; mass of a mouth's far asymptotic end \\
$\varepsilon$, $\varepsilon_2$ & amplitudes of the spherical and quadrupolar perturbations \\
$\tau$ & e-folding time of the unstable mode \\
$\theta_\pm$, $M_{\rm MS}$ & outgoing and ingoing null expansions; Misner--Sharp mass \\
$\Psi_4$, $E_{\rm GW}$ & Weyl scalar; radiated gravitational-wave energy \\
$E_\phi$ & radiated scalar-field energy \\
$L$, $\Delta x$ & side of the cubic domain; grid spacing \\
$\sigma_{\rm KO}$ & Kreiss--Oliger dissipation coefficient \\
\end{tabular}
\end{ruledtabular}
\end{table}

\subsection{Drainhole initial data}
\label{sec:model:drainhole}

Each body is a massive Ellis--Bronnikov drainhole~\cite{ellis1973,bronnikov1973}. In isotropic radius $r$ about the throat centre, with
\begin{equation}
X = \frac{r - a^2/4r}{a},\quad
\Omega = 1 + \frac{a^2}{4r^2},\quad
u = \frac{m}{a}\Bigl(\arctan X - \frac{\pi}{2}\Bigr),
\end{equation}
the exact static solution of Eq.~\eqref{eq:phantom} is
\begin{equation}
\alpha = e^{u},\quad
\gamma_{ij} = e^{-2u}\,\Omega^2\,\delta_{ij},\quad
\phi = C\arctan X,
\label{eq:drainhole}
\end{equation}
with $4\pi a^2C^2 = a^2+m^2$, extrinsic curvature $K_{ij}=0$, $\Pi=0$ and zero shift. The evolved variables are the conformal factor $\chi = e^{2u}\Omega^{-2}$ and the conformal metric $h_{ij}=\delta_{ij}$, with $\gamma_{ij}=h_{ij}/\chi$. Both constraints hold identically at $t=0$. Here $a$ is the throat scale and $m$ the ADM mass. The lapse, not the conformal factor, carries the mass, so $\chi$ at the throat stays in $[\clmChiThroatLo,\clmChiThroatHi]$ for $m/a\le1$ and mass costs no resolution. The minimal surface sits at $r_t=\tfrac12\bigl(m+\sqrt{m^2+a^2}\bigr)$, where $X=m/a$, and its areal radius is $R_{\min}=e^{-u(r_t)}\sqrt{a^2+m^2}=\sqrt{a^2+m^2}\,\exp\!\bigl[(m/a)\arctan(a/m)\bigr]$, which is $a$ at $m=0$. Our fiducial throat, $a=\clmThroatA$ and $m=\clmThroatM$, has the throat radius $R_\star\equiv R_{\min}=\sqrt5\,e^{\frac12\arctan2}=\clmRstar$ at $r_t=\tfrac12(1+\sqrt5)=\clmThroatCoordRadius$. The point $r\to0$ is the compactified spatial infinity ($\chi\sim r^4$) of the throat's far asymptotic end, the asymptotic region beyond the throat. At $m=0$ the throat is ultrastatic ($\alpha\equiv1$, $M_{\rm ADM}=0$).

\subsection{Binary initial data}
\label{sec:model:binary}

Two throats at separation $d$ (one-body radii $r_A$, $r_B$) start from their combined fields,
\begin{equation}
u = u_A + u_B,\qquad
\psi = 1 + \bigl(\sqrt{\Omega_A}-1\bigr) + \bigl(\sqrt{\Omega_B}-1\bigr),
\end{equation}
with $\chi = e^{2u}\psi^{-4}$, $\alpha = e^{u}$ and $\phi = \phi_A + \sigma\,\phi_B$, $\sigma=\pm1$; the constant tail of the $\arctan$ profiles is subtracted so that $\phi\to0$ at the boundary. The constraint solve of Sec.~\ref{sec:model:solve} turns this superposition into initial data. Momentum comes from the exact boost of Sec.~\ref{sec:model:boost}. The throats sit at $x=\mp d/2$ with momenta $P^i=(0,\pm p,0)$, tangential to the separation and opposite on the two throats, with $p$ in units of the one-body mass. The pair thus carries no net momentum and orbital angular momentum $J=pd$; $p=0$ is a release from rest. The relative sign $\sigma$ is the pair's one discrete parameter. Far from its throat each body is a mass $m$ carrying the scalar charge $q=Ca$ ($\phi_A\simeq-q/r_A$ once the constant tail is subtracted). A phantom mediator repels like charges and attracts opposite ones, so two throats at rest a large distance $d$ apart attract with
\begin{equation}
\begin{split}
F&=\frac{m^2}{d^2}\,\bigl(1-\sigma Q\bigr),\\
Q&\equiv\Bigl|\frac{F_\phi}{F_{\rm grav}}\Bigr|=\frac{4\pi q^2}{m^2}=\frac{a^2+m^2}{m^2}>1
\end{split}
\label{eq:force}
\end{equation}
always ($Q=\clmScalarQ$ for the fiducial throat): a like-signed pair repels and never merges under its own gravity, and $\sigma=\clmMergerSign$ is the merger channel.

\subsection{Constraint solve and matched data}
\label{sec:model:solve}

Superposed throats violate the Hamiltonian constraint and distort each other's mouths; we correct both. With $\Pi=0$, $K=0$ and $\gamma_{ij}=\Psi^4\delta_{ij}$, where $\Psi=e^{-u/2}\psi$ (so $\chi=\Psi^{-4}$), the Hamiltonian constraint $\mathcal H=0$ of Eq.~\eqref{eq:phantom} reads, in flat operators,
\begin{equation}
\nabla^2\Psi-\pi|\nabla\phi|^2\,\Psi+\tfrac18\hat A_{ij}\hat A^{ij}\,\Psi^{-7}=0 .
\label{eq:hamsolve}
\end{equation}
Here $\hat A_{ij}$ is the trace-free conformal extrinsic curvature. The single drainhole solves Eq.~\eqref{eq:hamsolve} exactly. For a pair we solve it for a correction $w$ to the starting $\Psi$ on the whole initial mesh hierarchy with AMReX's multigrid~\cite{amrex}, holding $\phi$, $\Pi$ and the lapse fixed (a Robin outer boundary); for a pair with momentum we solve the Hamiltonian and momentum constraints, $\mathcal H=0$ and $\mathcal M_i=0$, together (Sec.~\ref{sec:model:boost}).

Each mouth (the side of a throat facing our asymptotic region) is identified by the constants of its far asymptotic end. Near a throat's centre $\Psi=c/r+d_c+O(r)$, and the inversion $r'=c^2/r$ turns this into an asymptotically flat end with ADM mass and scalar charge
\begin{equation}
M_{\rm far}=2c\,d_c,\qquad Q_{\rm far}=\frac{4Cc^2}{a},
\label{eq:farside}
\end{equation}
both frozen in time, like a puncture's individual mass. The isolated drainhole has $M_{\rm far}=-m\,e^{\pi m/a}=\clmFarMassIso$ and $Q_{\rm far}=\sqrt{a^2+m^2}\,e^{\pi m/a}/\sqrt{4\pi}=\clmFarChargeIso$. A pair $(M_{\rm far},Q_{\rm far})$ names one isolated drainhole, and its near-side mass $m'$ is the mouth's one-body mass.

A second throat in the slice rescales each mouth: near $A$ the companion's constants multiply the throat's own field, $\Psi\to e^{-u_B(x_A)/2}\,\Psi$. Because $\phi$ fixes a static throat's coordinate size, the puncture coefficient $c$ alone cannot undo this. We call data \emph{matched} when each mouth's far-end mass and charge equal the isolated throat's. The matched placement gives each throat a coordinate scale $\lambda$ ($a\to\lambda a$ and $m\to\lambda m$ in $\phi$, $u$ and $\Omega$; $C$ is scale-free), sets $\lambda=(c/c_{\rm iso})^2$ with $c_{\rm iso}=\tfrac a2e^{\pi m/2a}$, which holds $Q_{\rm far}$ at its isolated value, and iterates $c$ until $M_{\rm far}$ is isolated too. Each mouth is then locally the isolated static drainhole (Table~\ref{tab:placement}): its $R_{\min}$, which the solve does not constrain, lies within $\clmMatchedRminDev\%$ of $R_\star$, and $m'=1$. With momentum the iteration converges the same way.

\begin{table}[t]
\caption{Each mouth of the matched $d=8$ pair at $t=0$ ($L=64$, level 3) against the isolated throat: minimal areal radius, far-end mass [Eq.~\eqref{eq:farside}], one-body mass $m'$, and the throat's own $M_{\rm ADM}$ from Eq.~\eqref{eq:madmvol}. The pair's $M_{\rm ADM}$ is $\clmHeadonCsmAdmMass$ on the fiducial $L=128$ grid.}
\label{tab:placement}
\begin{ruledtabular}
\begin{tabular}{lcccc}
data & $R_{\min}$ & $M_{\rm far}$ & $m'$ & $M_{\rm ADM}$ \\
\hline
isolated throat & \tcell{Riso}{\clmPlaceTabRiso} & \tcell{Miso}{$\clmPlaceTabMiso$} & 1 & \tcell{Aiso}{\clmPlaceTabAiso} \\
matched pair, each mouth & \tcell{Rmat}{\clmPlaceTabRmat} & \tcell{Mmat}{$\clmPlaceTabMmat$} & 1 & \tcell{Amat}{\clmPlaceTabAmat} \\
\end{tabular}
\end{ruledtabular}
\end{table}

The pair's ADM mass follows from Gauss's theorem on Eq.~\eqref{eq:hamsolve}, with $\Psi\to1+M_{\rm ADM}/2r$ and punctures contributing $-4\pi c_X\delta_X$:
\begin{equation}
M_{\rm ADM}=2\sum_X c_X-\frac1{2\pi}\int\Bigl(\pi|\nabla\phi|^2\Psi-\tfrac18\hat A_{ij}\hat A^{ij}\Psi^{-7}\Bigr)d^3x ,
\label{eq:madmvol}
\end{equation}
which returns the exact throat's $M=1$ to $\clmVolumeIdentityExact\%$ and, unlike a fit to the monopole tail, does not read the outer boundary's offset of $w$ as mass. For matched pairs at rest (Table~\ref{tab:energy}) the excess over two isolated throats follows
\begin{equation}
E_b\equiv M_{\rm ADM}-2m=-\lambda^2\,\frac{m^2}{d}\,\bigl(1+\sigma Q\bigr)
\label{eq:ebind}
\end{equation}
for both signs, to $\clmEnergyLawDevNear\%$ at $d=8$. It is the Newtonian binding $-\lambda^2m^2/d$ plus the phantom cross energy $-\sigma\!\int\nabla\phi_A\!\cdot\!\nabla\phi_B\,d^3x=-\sigma\lambda^2(a^2+m^2)/d$ of two charges $\lambda q$, with $\lambda\to1$ as $d\to\infty$; the solve adds nothing else. In the merger channel $\sigma=-1$ the positive scalar term dominates, so the attracting pair weighs more than its parts ($E_b\,d\to\clmEnergyLimitAttract$) and the repelling pair less ($E_b\,d\to\clmEnergyLimitRepel$). Equation~\eqref{eq:ebind} is the ADM mass of data at fixed far-end constants, not the two-body potential; read as one, it gives the wrong sign in both channels. Equation~\eqref{eq:force} is the standard force for a phantom mediator, which reverses the sign of ordinary scalar exchange; Sec.~\ref{sec:pair:sign} tests it.

\begin{table}[t]
\caption{Mass of matched pairs at rest [Eq.~\eqref{eq:madmvol}] against two isolated throats on the same grid ($L=128$, $\Delta x=0.0625$ at the throats), and Eq.~\eqref{eq:ebind}. $\lambda$: each throat's coordinate scale.}
\label{tab:energy}
\begin{ruledtabular}
\begin{tabular}{r|ccc|ccc}
 & \multicolumn{3}{c|}{$\sigma=-1$ (attract)} & \multicolumn{3}{c}{$\sigma=+1$ (repel)} \\
$d$ & $\lambda$ & $E_b$ & Eq.~\eqref{eq:ebind} & $\lambda$ & $E_b$ & Eq.~\eqref{eq:ebind} \\
\hline
$8$ & \tcell{Lo8}{\clmEnergyLambdaAttractEight} & \tcell{Eo8}{$\clmEnergyBindAttractEight$} & \tcell{Po8}{$\clmEnergyLawAttractEight$} & \tcell{Ll8}{\clmEnergyLambdaRepelEight} & \tcell{El8}{$\clmEnergyBindRepelEight$} & \tcell{Pl8}{$\clmEnergyLawRepelEight$} \\
$12$ & \tcell{Lo12}{\clmEnergyLambdaAttractTwelve} & \tcell{Eo12}{$\clmEnergyBindAttractTwelve$} & \tcell{Po12}{$\clmEnergyLawAttractTwelve$} & \tcell{Ll12}{\clmEnergyLambdaRepelTwelve} & \tcell{El12}{$\clmEnergyBindRepelTwelve$} & \tcell{Pl12}{$\clmEnergyLawRepelTwelve$} \\
$16$ & \tcell{Lo16}{\clmEnergyLambdaAttractSixteen} & \tcell{Eo16}{$\clmEnergyBindAttractSixteen$} & \tcell{Po16}{$\clmEnergyLawAttractSixteen$} & \tcell{Ll16}{\clmEnergyLambdaRepelSixteen} & \tcell{El16}{$\clmEnergyBindRepelSixteen$} & \tcell{Pl16}{$\clmEnergyLawRepelSixteen$} \\
$24$ & \tcell{Lo24}{\clmEnergyLambdaAttractTwentyFour} & \tcell{Eo24}{$\clmEnergyBindAttractTwentyFour$} & \tcell{Po24}{$\clmEnergyLawAttractTwentyFour$} & \tcell{Ll24}{\clmEnergyLambdaRepelTwentyFour} & \tcell{El24}{$\clmEnergyBindRepelTwentyFour$} & \tcell{Pl24}{$\clmEnergyLawRepelTwentyFour$} \\
$48$ & \tcell{Lo48}{\clmEnergyLambdaAttractFortyEight} & \tcell{Eo48}{$\clmEnergyBindAttractFortyEight$} & \tcell{Po48}{$\clmEnergyLawAttractFortyEight$} & \tcell{Ll48}{\clmEnergyLambdaRepelFortyEight} & \tcell{El48}{$\clmEnergyBindRepelFortyEight$} & \tcell{Pl48}{$\clmEnergyLawRepelFortyEight$} \\
\end{tabular}
\end{ruledtabular}
\end{table}

Every binary starts from the matched, solved data above, so each mouth has one-body mass $m'=1$ to $\clmMatchedOneBodyTol$. A quoted momentum $p$ is therefore per one-body mass, with boost speed $v=p/\sqrt{1+p^2}$. The pair's mass is the $M_{\rm ADM}$ of Eq.~\eqref{eq:madmvol} at rest and of Sec.~\ref{sec:model:boost} in motion.

\subsection{Boosted throats}
\label{sec:model:boost}

\begin{figure*}[t]
\centering
\includegraphics[width=0.80\textwidth]{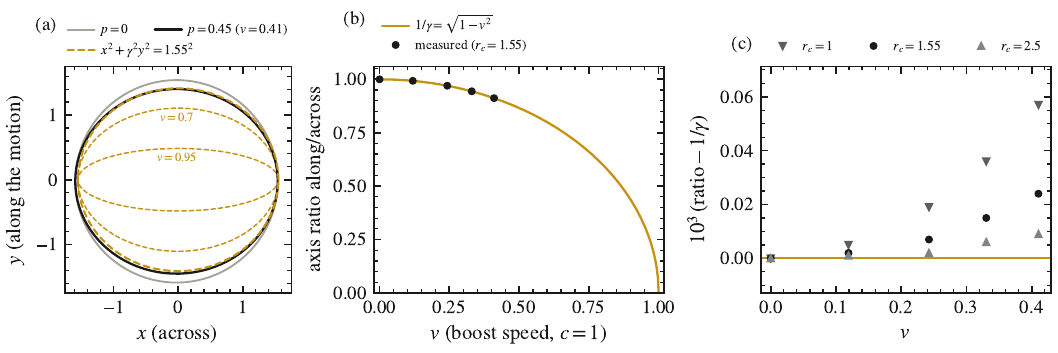}
\caption{Shape of the boosted throat at $t=0$: $\chi$ contours and their axis ratio against boost speed $v$. (a)~The $\chi$ contour just inside the throat, at rest and at $p=\clmSpiralMomentumFlyby$, the latter on its analytic ellipse; fainter ellipses continue the law to $v=0.7$ and $0.95$. (b)~Axis ratio (along/across the motion) against $v$, on the curve $1/\gamma=\sqrt{1-v^2}$. (c)~Residual from $1/\gamma$ at three contour levels. The boosted data carry the special-relativistic shape, and both constraints hold at the discretisation error of the exact solution.}
\label{fig:boost_contraction}
\end{figure*}

A throat with ADM momentum $p$ is the exact drainhole boosted to $v=p/\sqrt{m^2+p^2}$ and cut at $t=0$. We boost the metric, $K_{ij}$, $\phi$ and $\Pi$ together, with the boosted lapse and shift, so the slice is an exact solution of both constraints and the scalar moves with its throat ($\Pi\neq0$). On the slice every $\chi$ contour, the throat's included, is an ellipsoid contracted along the motion by $1/\gamma=m/\sqrt{m^2+p^2}$, with $\gamma$ the Lorentz factor, as a boosted black hole's horizon is. The measured axis ratios follow this law to $\clmBoostAxisResidual$ at every $p$ up to $\clmSpiralMomentumFlyby$ (Fig.~\ref{fig:boost_contraction}).

A pair is built from two boosted throats as in Sec.~\ref{sec:model:solve}, with the companion's $K_{ij}$ and $\Pi$ tapered to zero inside each throat. The solve then corrects both constraints, for $w$ and a vector potential $W$ ($\hat A_{ij}\to\hat A_{ij}+(\mathbb{L}W)_{ij}$, with $\mathbb{L}$ the conformal Killing operator). The taper acts only inside each throat, near its compactified infinity, so no unsolved defect is left to radiate (Sec.~\ref{sec:gw}). Three tests validate the boosted solve. On an isolated boosted throat on the fiducial grid the solve leaves the data unchanged (puncture correction $|w_0|=\clmBoostSolveW$). On the pair at rest it reproduces the scalar solve's $c$, $\sigma$ and $w_0$ to $\clmBoostRestRepro$. On the opposite-signed $d=12$, $p=\pm\clmSpiralMomentumTwentyFive$ pair, each far asymptotic end comes out the isolated throat's to $\clmBoostPairFar$. Equation~\eqref{eq:madmvol} needs conformally flat data with $K=\Pi=0$, so a moving pair's mass is the boosted background's ADM energy (each throat's $\gamma\lambda_X m$ and its puncture shift) plus the monopole of $w$ on the box faces. Where both apply, the face value reads $\clmBoostFaceBias\%$ below Eq.~\eqref{eq:madmvol}. For the moving $d=12$ pairs, a full ADM surface integral of the $t=0$ slice lies $\clmMassTzeroExcessLo$--$\clmMassTzeroExcessHi\%$ above this prescription, and we quote their energies per that measured mass.

\subsection{Initial perturbations}
\label{sec:model:seeds}

We probe the instability with added perturbations of the conformal factor only, a Gaussian shell on the minimal surface,
\begin{equation}
\psi \to \psi\bigl[1 + \varepsilon\, g(r)\bigr],\qquad
g = e^{-(r-r_t)^2/s^2},\quad s = a/4,
\label{eq:seed}
\end{equation}
which changes the throat's areal radius by a fraction ${\approx}\clmSeedArealFactor\varepsilon$, and a quadrupolar factor
\begin{equation}
\psi \to \psi\bigl[1 + \varepsilon_2\, g(r)\,P_2(\cos\theta)\bigr]
\end{equation}
about the $z$ axis through the throat. Neither perturbation is constraint-solved. $K_{ij}$ and $\Pi$ stay untouched, so the momentum constraint holds at $t=0$. The drainhole, however, solves the linear equation $\nabla^2\Psi=\pi|\nabla\phi|^2\Psi$, so the spherical perturbation leaves the defect $\mathcal H=-8\varepsilon\,\Psi^{-4}(1+\varepsilon g)^{-5}(\nabla^2g+2\nabla\ln\Psi\cdot\nabla g)$. The defect is confined to the narrow shell and large there, with a peak of $\clmSeedDefectPeak$--$\clmSeedDefectPeakHi\times16\pi|\rho|$ at $\varepsilon=\pm10^{-2}$. The quadrupolar perturbation's defect is pure $\ell=2$ at first order. CCZ4, the conformal covariant Z4 formulation~\cite{alic2012}, feeds the defect into its constraint variable $\Theta$ and, within the first steps, into $\mathcal M_i$ (Fig.~\ref{fig:constraints}(b)). The constraint defect does not select the branch. The same $\varepsilon=+0.1$ perturbation, carried through the constraint solve (Sec.~\ref{sec:model:solve}) as a larger puncture coefficient $c=1.1\,c_{\rm iso}$, is trapped already on its $t=0$ slice. We encode it in $c$ because the solve, unique at fixed $\phi$, would remove a perturbation of the conformal factor. The unsolved $+0.1$ perturbation reaches the same branch at $t=\clmCeilingPlusMotsTime$ (Sec.~\ref{sec:single:branching}). Above $\varepsilon\simeq\clmSeedBagThreshold$ the spherical perturbation turns the minimal surface into a maximum between two minima. The spherical perturbation selects the branch but cannot radiate, since a spherical collapse has no $\ell\ge2$ moment. The quadrupolar perturbation, with no $\ell=0$ part at first order, leaves the branch to the spherical one and gives a collapsing throat a quadrupole to shed, an isolated throat's only gravitational-wave signal (Sec.~\ref{sec:single:gw}).

\section{Numerical setup}
\label{sec:setup}

This section gives the evolution scheme, the grid and the diagnostics; Table~\ref{tab:matrix} summarises our simulations.

\begin{table*}[t]
\caption{Summary of the $\clmDetRunsTotal$ simulations, in groups that each vary one parameter against a named reference or repeat one run; the pairs start on constraint-solved, boosted data (Sec.~\ref{sec:model:boost}). Left: the single throat and the initial-data construction. Right: the binaries and their controls. Unless stated, $a=\clmThroatA$, $m=\clmThroatM$ ($R_{\min}=\clmRstarShort$), $\sigma=\clmMergerSign$, $d=\clmPairSeparation$ and $L=\clmBoxL$ ($\clmBigBoxL$ for binaries with more than three refinement levels). The data release lists every run.}
\label{tab:matrix}
\scriptsize
\begin{ruledtabular}
\begin{tabular}{lp{0.215\textwidth}cc|lp{0.195\textwidth}cc}
group & parameters varied & runs & Sec. & group & parameters varied & runs & Sec. \\
\hline
single throat & levels \clmDetKnobLoneLevelsLo--\clmDetKnobLoneLevelsHi; $\chi$ floor; $\Delta t$; $L{=}\clmDetKnobLoneL$; $m{=}\clmDetKnobLoneM$ & \clmDetRunsLoneThroat{} & \ref{sec:single} & two throats at rest & $\sigma$; $a=\clmDetKnobMatchedALo$--$\clmDetKnobMatchedAHi$; $d=\clmDetKnobMatchedDLo$--$\clmDetKnobMatchedDHi$; one like-sign pair to $t{=}100$ & \clmDetRunsMatchedRest{} & \ref{sec:pair} \\
spherical perturbations & $\varepsilon=\pm\clmDetKnobKickEpsLo\ldots\pm\clmDetKnobKickEpsHi$; levels \clmDetKnobKickLevelLo--\clmDetKnobKickLevel{} & \clmDetRunsSphericalKicks{} & \ref{sec:single:branching} & head-on, $d{=}8$ & restart segments, levels \clmDetKnobHoMotsLevelLo--\clmDetKnobHoMotsLevelHi; post-processing MOTS search;\newline first-law window & \clmDetRunsHeadonModeThree{} & \ref{sec:headon} \\
quadrupolar perturbations & $\varepsilon_2=\clmDetKnobQuadLo$--$\clmDetKnobQuadHi$; unperturbed run & \clmDetRunsQuadKicks{} & \ref{sec:single:gw} & $d{=}6$ merger & $p{=}\clmDetKnobChainPTang$ tangential; levels \clmDetKnobChainLevelLo--\clmDetKnobChainLevelHi;\newline $\sigma_{\rm KO}$; $\chi$ floor; final restart segment & \clmDetRunsMergerChain{} & \ref{sec:spiral:inspiral} \\
reruns with scalar extraction & unperturbed, level \clmDetKnobScalarLevel, $L{=}\clmDetKnobScalarLLo$/$\clmDetKnobScalarLHi$;\newline $\varepsilon{=}{+}\clmDetKnobScalarEps$ & \clmDetRunsScalarReruns{} & \ref{sec:gw:censorship} & $d{=}12$ runs & $p=\clmDetKnobDTwelvePLo$ tangential, $\clmDetKnobDTwelvePHi$ inward & \clmDetRunsDTwelveArms{} & \ref{sec:spiral:race} \\
boosted throat & $p=0$--$\clmDetKnobBoostPHi$ at $t{=}0$; paired collapse runs,\newline levels 3/4 & \clmDetRunsBoostedThroat{} & \ref{sec:model:boost} & fly-by and $E_{\rm GW}$ scan & $p=\clmDetKnobEgwPLo$--$\clmDetKnobEgwPHi$; $\chi$-floor and\newline level-5 extensions & \clmDetRunsEgwScan{} & \ref{sec:gw:flyby} \\
solver tests & boosted pairs at $t{=}0$ on the fiducial grid; the pair at level 3 to $t{=}20$; a perturbed single throat through the solve & \clmDetRunsSolveVerifications{} & \ref{sec:model:boost} & vacuum binary black hole & same $d$, $p$, masses; no scalar, no sponge & \clmDetRunsVacuumBBH{} & \ref{sec:controls:bbh} \\
placement & $d=\clmDetKnobPlaceCsmDLo$--$\clmDetKnobPlaceCsmDHi$, one step each & \clmDetRunsMatchedPlacement{} & \ref{sec:model:solve} & convergence & head-on at three resolutions, $L{=}\clmFarzoneBoxL$; $\sigma_{\rm KO}$;\newline $d{=}12$ run at level \clmDetKnobConvArmLevel, wave zone refined & \clmDetRunsConvergence{} & \ref{app:health} \\
\end{tabular}
\end{ruledtabular}
\end{table*}

\subsection{Evolution scheme}
\label{sec:setup:scheme}

We evolved the CCZ4 system~\cite{alic2012} with constraint-damping parameters $(\kappa_1,\kappa_2,\kappa_3)=(\clmCczKappaOne,0,\clmCczKappaThree)$, $1{+}\log$ slicing~\cite{bona1995} and a Gamma-driver shift~\cite{alcubierre2003gauge,campanelli2006,baker2006} with $\eta=\clmShiftEta$, both started from the initial data's lapse and shift. We used fourth-order stencils, RK4 time integration and Kreiss--Oliger (KO) dissipation with coefficient $\sigma_{\rm KO}=\clmKOsigma$. The time step $\Delta t=\clmTimeStepFactor\,\Delta x$, an order of magnitude below the usual CCZ4 Courant factor, keeps the evolution stable. Every resolution statement below is at fixed $\Delta t/\Delta x$; Appendix~\ref{app:health} tests convergence in resolution and in extraction radius.

\subsection{Grid, boundaries and regularisation}
\label{sec:setup:grid}

The domain is a cube of side $L=\clmBoxL$ centred on the system, with a $\clmBaseN^3$ base grid ($\Delta x=\clmDxBase$). Nested boxes, halving $\Delta x$ per level and centred on the tracked throats (half-width $\clmFinestHalfWidth$ per throat at the finest level of the main runs), are redrawn every $\clmRegridSteps$ steps. The main runs used \clmProdLevels{} refinement levels (finest $\Delta x=\clmFinestDx$), higher-resolution runs up to level $\clmResArmsLevelHi$. Every binary with more than three refinement levels, and two single-throat runs, use $L=\clmBigBoxL$: the same cells in a box twice as wide. Level $j$ thus means a finest cell $\Delta x=\clmDxBase/2^j$. The one exception, the inflating throat (Sec.~\ref{sec:single:fate}), is evolved in an octant of an $L=\clmFourBoxL$ box whose \clmFourLevel{} levels give $\Delta x=\clmFourFinestDx$, the resolution of our level-3 runs. Outer boundaries are Sommerfeld, backed by a spherical sponge over the outer quarter of the half-width, in which KO dissipation ramps to $\sigma_{\rm sp}=\clmSpongeStrength$ as the \clmSpongePower{} power of the depth. Every extraction sphere lies inside the sponge's inner radius. The sponge damps grid-scale noise but is transparent to long-wavelength gauge and physical waves, which reflect from the box faces (Sec.~\ref{sec:headon:wall}). The vacuum controls (Sec.~\ref{sec:controls:bbh}) instead run \texttt{GRTeclyn}'s binary-black-hole example with its own numerical settings and no sponge.

Each throat carries a compactified second infinity at its centre, where $\chi\sim r^4$, so the evolved variables are floored at $\chi\ge\clmChiFloor$ and $\alpha\ge\clmLapseFloor$. A variant (V) floors $\chi$ at $\clmChiFloorVariant$ and applies $\clmChiRhsFloor$ only where $1/\chi$ is evaluated. Runs that differ only in the $\chi$ floor agree to \clmControlChiDigits{} digits after touching it. In the runs Table~\ref{tab:mods} marks O, the $1{+}\log$ source is off within coordinate radius $\clmFreezeRadiusFull$ of each tracked throat centre (the compactified far asymptotic end) and restored by $\clmFreezeRadiusStart$ ($r_t=\clmThroatCoordRadius$), so that the moving puncture does not drive the slicing. In the runs it marks D, the scalar is damped exponentially wherever the lapse has collapsed below $\clmDampLapseStart$: inside merged cores, and between fly-by mouths as they approach. The $p=\clmSpiralMomentumFlyby$ fly-by, rerun without damping to $t=70$, matches the Weyl scalar $\Psi_4$ of its damped counterpart to $\clmDampOffPsi$ of the peak. The damping first acts at $t=\clmDampOnTime$, so this comparison covers only $R\lesssim\clmDampOffReach$: the burst peak, not the outer sphere the energy is read from.

\begin{table}[t]
\caption{Numerical modifications per run beyond the defaults of Sec.~\ref{sec:setup:scheme} ($\sigma_{\rm KO}=\clmKOsigma$, $\chi\ge\clmChiFloor$, $1{+}\log$ everywhere, no damping). Restart segments follow semicolons, with their start times in parentheses. V is the variant $\chi$ floor; ``2nd run'' marks a second run that differs only in that setting. Last column: $1{+}\log$ off near the throats (O), scalar damped (D).}
\label{tab:mods}
\scriptsize
\setlength{\tabcolsep}{2pt}
\begin{ruledtabular}
\begin{tabular}{llll}
Run & Levels & $\chi$ floor; $\sigma_{\rm KO}$ & O, D \\
\hline
Single throat & 2--4 & V (2nd run) & -- \\
Moving throat & 3, 4 & -- & O \\
Rest pairs & 3 & -- & -- \\
Head-on & 5; 6 (35); 4 (50) & V; 0.3 (2nd run) & -- \\
Head-on, $L=\clmFarzoneBoxL$ & 5, 6, 7 & $10^{-4}$; 0.3 & -- \\
$d=6$ merger & 5; 5 (25, 30); 4 (35) & $10^{-4}$ (30); 1.0 (25) & O, D \\
$p=\clmSpiralMomentumProd$ & 4, 5 & -- & -- \\
$p=\clmSpiralMomentumTwentyFive$ & 5 & -- & O, D \\
$p=\clmSpiralMomentumFlyby$ & 4 & -- & O, D \\
$p=\clmEgwPSixty$ & 4; 4 (40, 50); 5 (40, 60) & $10^{-4}$ (40) & O, D \\
$p=\clmEgwPNinety$ & 4; 4, 5 (40) & $10^{-4}$ (40) & O, D \\
\end{tabular}
\end{ruledtabular}
\end{table}

\subsection{Diagnostics}
\label{sec:setup:diagnostics}

Every coarse step the code writes the volume-averaged norms $\|\mathcal H\|=(\int\mathcal H^2\,\mathrm{d}V/L^3)^{1/2}$ and $\|\mathcal M\|=(\int\mathcal M_i\mathcal M_i\,\mathrm{d}V/L^3)^{1/2}$, evaluated on the base grid over the whole domain after the finer levels are restricted onto it. They do not resolve the throats, so they monitor the evolution rather than local accuracy and compare only within one box. At $t=0$ they are set by the base grid's stencils on the throat profiles, not by the data. We quote a run only while it is reliable: before its constraint norms begin to grow steeply (on the fly-bys, before $\|\mathcal H\|$ first exceeds $2.5\times10^{-2}$) and before grid-scale noise appears in $K$ and $\Psi_4$. We say the run is reliable up to that time.

Both throat centres are tracked, and about each the areal radius $R(r)$ is scanned on coordinate spheres out to a fixed edge. Its minimum is the throat; an inflating throat that outgrows the scan is bounded from above by the edge reading. On the same spheres the null expansions $\theta_\pm$ are computed with the outward direction fixed by increasing areal radius, so that a dynamic throat is marginally anti-trapped~\cite{hochberg_visser1998,hayward1999}. An in-code monitor that orients by coordinate radius, the wrong way beyond a throat, serves only for live checks.

We call \emph{the round-sphere scan} the search, about each centre, for the outermost coordinate sphere that is trapped everywhere ($\max\theta_+=0$, $\theta_-<0$). A trapped sphere lies inside the apparent horizon, so this sphere bounds the marginally outer trapped surface (MOTS, Sec.~\ref{sec:setup:mots}) from inside. On its spheres, as on the MOTS finder's surfaces, we evaluate the areal radius $R$ and the Hawking mass~\cite{hawking1968}, the Misner--Sharp mass~\cite{misner_sharp1964} on round spheres,
\begin{equation}
\frac{2M_{\rm MS}}{R} \;=\; 1 + \frac{R^{2}\,\langle\theta_+\theta_-\rangle}{4},
\label{eq:misner_sharp}
\end{equation}
with the product averaged over the sphere and $\theta_\pm$ computed for the null normals $k^a=n^a+s^a$ and $l^a=n^a-s^a$ ($n^a$ the slice's unit normal, $s^a$ the sphere's outward unit normal, so $k\cdot l=-2$). On an exact MOTS $M_{\rm MS}=R/2$; on the scan's spheres $M_{\rm MS}$ exceeds $R/2$ by at most $\clmShapeSystLone\%$ about a single collapsed throat, which stays nearly round.

Radiation is read from $\Psi_4$ on three spheres from $R=\clmExtractRadiusTen$ to $\clmExtractRadiusEighteen$ on the single throats (a fourth on their wave runs), and on four from $R=\clmExtractRadiusTwenty$ to $\clmExtractRadiusFortyFour$ on every $L=\clmBigBoxL$ binary, the head-on adding the inner three. It is decomposed in spin-weight $-2$ harmonics about $z$, the orbital axis. The head-ons fall along $x$, so their quadrupole $h'_{20}$ about the collision axis splits in the $z$-based modes: $(2,0)=-h'_{20}/2$ carries a quarter of its power and $(2,\pm2)$ the rest ($|h_{22}/h_{20}|=\sqrt{3/2}$ to $\clmHeadonAxisRatioDev\%$ on every sphere). We therefore quote their $(2,0)$ as $h'_{20}$. Every amplitude quoted is $r\Psi_4$, and strain is its fixed-frequency integral~\cite{reisswig2011}. On the head-on the in-code and post-processing extractions agree at the burst peak to $\clmPsiPipelines\%$. On the single throats the in-code extraction is dominated by its grid-scale floor, and only the post-processing one is used. The scalar field is extracted on the single throats' spheres and, on the binaries, at $R=\clmGwScalarRadiusInner$, $20$, $\clmGwScalarRadiusOuter$ and $44$ (the head-on adding $10$ and $18$). It yields $\phi$ and $\Pi$ in spin-weight-$0$ harmonics, $A_{\ell m}=R\,\phi_{\ell m}$, and the kinematic flux $F_{\rm kin}=-R^2\oint\Pi\,\partial_r\phi\,\mathrm{d}\Omega$, positive for an outgoing canonical wave (Sec.~\ref{sec:gw:scalar}). An optional diagnostic records the cell and the one-step history of any floating-point failure.

\subsection{Marginally outer trapped surfaces}
\label{sec:setup:mots}

A MOTS is a closed two-surface of the slice on which the outgoing null expansion vanishes, $\theta_+=0$, while
the ingoing one is negative, $\theta_-<0$; the outermost MOTS is the apparent horizon. Unlike the event
horizon, a MOTS is read off a single slice. Since the phantom scalar violates the null energy condition, a MOTS
need not lie inside an event horizon. Every ``horizon'' and ``censored'' in this paper refers to MOTSs on the
evolved slices. ``Black hole'' names the region inside such a dynamical apparent horizon, whose area this matter can shrink (Sec.~\ref{sec:single:collapse}), not a stationary black hole, which the remnants only approach (Sec.~\ref{sec:headon:wall}).

The MOTS finder runs in post-processing on every saved slice, so horizon coverage matches the output interval. It
expands the surface about the tracked centre as $r=h(\theta,\phi)$ in real spherical harmonics up to degree
$\ell_{\max}=\clmFlowFinderLmax$ on the head-on slices and $\ell_{\max}=\clmFlowFinderLmaxAlt$ on the orbital
ones, and solves $\theta_+[h]=0$ by Newton iteration on metric and extrinsic-curvature data interpolated from
level 3 on the head-on and level 2 on the orbits, whatever the run's finest level, with the areal orientation of
Sec.~\ref{sec:setup:diagnostics}. The first surface of a run is found by the damped fast
flow~\cite{gundlach1998}; each later slice warm-starts from the previous surface and converges in a few
iterations, until every projected harmonic of $\theta_+$ is below $10^{-6}$. From the converged surface
we quote the areal radius $R=\sqrt{A/4\pi}$, the Misner--Sharp mass of Eq.~\eqref{eq:misner_sharp}, and the
harmonic content beyond $\ell=0$ as the deformation. The flow method recovers Kerr--Schild Schwarzschild
($R=2M$ to $\clmFlowFinderSchw\%$) and finds no trapped surface outside the time-symmetric Ellis throat. Every
binary horizon number in this paper (formation times, $R$, $M_{\rm MS}$) is read from these surfaces. The single
throats' nearly round collapse horizons come from the round-sphere scan (within $\clmShapeSystLone\%$,
Sec.~\ref{sec:setup:diagnostics}). On a binary the scan under-reads a forming, deformed horizon or misses it, and
serves only as a live monitor.

\section{The single wormhole}
\label{sec:single}

This section asks how an isolated massive throat leaves its static state, and how it ends.

\begin{figure*}[t]
\centering
\includegraphics[width=0.80\textwidth]{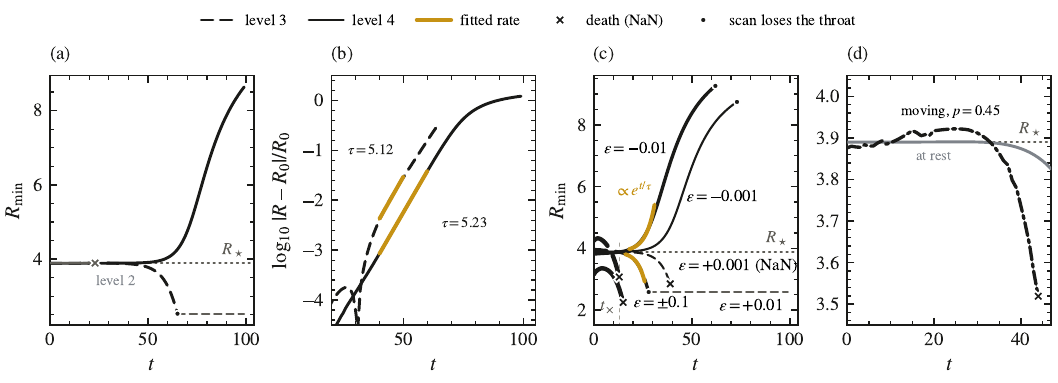}
\caption{Minimal areal radius $R_{\min}$ of the isolated throat against time, from exact static data. (a)~Levels 2--4 without an added perturbation: level 3 (dashed), level 4 (solid) and level 2 (grey, failing first at the cross); dotted, $R_\star$. Dots mark where the radius scan loses the throat; muted lines continue its last reading. (b)~$|R-R_0|/R_0$ at levels 3 and 4, with small-amplitude exponential fits (gold). (c)~Level 3 with added perturbations, Eq.~\eqref{eq:seed}: $\varepsilon=\pm10^{-2}$ (heavy), $\pm10^{-3}$ (light) and $\pm0.1$ (heaviest); dashed, outward ($\varepsilon>0$); gold, local exponential fits; crosses, failures at the compactified origin. Constraint norms: Fig.~\ref{fig:constraints}(a,b). (d)~The Lorentz-boosted throat (dash-dot, $p=\clmSpiralMomentumFlyby$, level 3) against the resting level-3 run (grey). The throat leaves its static state on one exponential mode, the sign of the initial perturbation selects collapse or inflation, and uniform motion changes neither.}
\label{fig:single_throat}
\end{figure*}

\subsection{The static throat as a fixed point}
\label{sec:single:fixedpoint}

Without an added perturbation, the exact drainhole stays static until truncation error excites its unstable mode. At level 3 the throat stays at $R_\star$ to $\clmHoldTolLevelThree\%$ for $\clmHoldTimeLevelThree$ time units and to $\clmHoldTolOnePct\%$ until $t\simeq\clmHoldTimeOnePct$ (Fig.~\ref{fig:single_throat}(a)); at level 4 it is exact to \clmLevelFourDigits{} digits at $t=30$. Each halving of the cell delays the departure by $\clmOnsetShiftLo$--$\clmOnsetShiftHi$ units, as expected when a mode of fixed growth rate starts from a smaller truncation error. Level 2 fails at the origin at $t=\clmLevelTwoDeath$, before the throat departs. The static phase scales with the mass: at level 3 the $m=\clmHalfMassM$ throat stays within $\clmHalfMassHoldTol\%$ only to $t\approx\clmHalfMassHoldTime$, against the fiducial $m=\clmThroatM$ throat above, and fails at the origin at $t=\clmHalfMassDeath$.

\subsection{The unstable mode}
\label{sec:single:instability}

The departure is a single exponential mode (Fig.~\ref{fig:single_throat}(a,b)). Fits below $\clmTauFitDev\%$ deviation give the e-folding time $\tau=\clmTauLevelThree$ (level 3) and $\clmTauLevelFour$ (level 4). In units of the throat's light-crossing time, the proper-time e-folding time is $T\equiv\tau\,\alpha_{\rm th}/R_\star=\clmProperEfoldThree$ and $\clmProperEfoldFour$, where $\alpha_{\rm th}=e^{u(r_t)}=\clmThroatLapse$ is the static lapse at the throat. Our drainhole is the $\gamma_1=m/a=\clmThroatGammaOne$ member of the static family of Gonz\'alez, Guzm\'an and Sarbach (their metric with $b=a$, $\gamma_0=-\pi\gamma_1/2$), for which their linear analysis gives $T=\clmGgsMatched$~\cite{gonzalez2009lin}. We solved the linearised field equations of this solution ourselves, with their regularised master equation and, independently, in the gauge $\delta\phi=0$; both give $T=\clmLinT$, i.e.\ $\tau_{\rm lin}=\clmLinTau\,M$. The measured e-folding time agrees with it to $\clmGgsAgreement\%$ at level 3 and $\clmGgsAgreementFour\%$ at level 4; larger departures grow more slowly. The initial perturbation is truncation error, and its sign flips with resolution: level 3 collapses and level 4 inflates. These are the two branches found in spherically symmetric evolutions of the massless~\cite{shinkai2002,calhoun2022} and massive~\cite{gonzalez2009,xu2025} throat and in 3D for the massless one~\cite{shirokov2026}; here they are reached from exact massive data in 3D.

\subsection{Branching: collapse versus inflation}
\label{sec:single:branching}

An added perturbation, Eq.~\eqref{eq:seed}, overrides the truncation error (Fig.~\ref{fig:single_throat}(c)). At $\varepsilon=\pm10^{-2}$ and $\pm10^{-3}$ an outward perturbation ($\varepsilon>0$) always collapses and an inward one always inflates. The radii of each $\pm\varepsilon$ pair cross at the $\varepsilon$-independent $t_\times\simeq\clmCrossingMean$. A tenfold smaller perturbation reaches equal horizon radius $\clmMatchedDelayLo$--$\clmMatchedDelayHi$ units later, as linear response on one mode predicts. The branching survives refinement: at level 4, $+10^{-2}$ still forms a horizon at $t=\clmHorizonTimeLevelFour$ and $-10^{-2}$ still inflates, the pair crossing at $t=\clmCrossingLevelFour$ against level 3's $\clmCrossingLevelThree$. The amplitude also sets the collapse time: horizons form at $t=\clmHorizonTimeEpsTwo$ ($\varepsilon=+10^{-2}$) and $t=\clmHorizonTimeEpsThree$ ($+10^{-3}$), against $t\le\clmHorizonTimeNoise$ for the unperturbed throat (its first output slice after the departure already contains a MOTS). At $\varepsilon=\pm0.1$ both runs collapse. The $+0.1$ run is trapped from $t=\clmCeilingPlusMotsTime$; the $-0.1$ run re-expands, then collapses, unlike the smaller inward perturbations, and is trapped from $t=\clmCeilingMinusMotsTime$. Both later fail at the compactified origin with a NaN (a floating-point failure). Linear response holds up to $|\varepsilon|=10^{-2}$.

\subsection{The throat in motion}
\label{sec:single:moving}

A uniformly moving throat keeps the resting throat's branch and rate. The Lorentz-boosted $p=\clmSpiralMomentumFlyby$ throat (Sec.~\ref{sec:model:boost}; level 3, $v=\clmBoostSpeed$) stays at $R_\star$ until it falls $1\%$ below its start at $t=\clmBoostMovingOnePct$, against the resting throat's $t\simeq\clmHoldTimeOnePct$. It then collapses on the resting throat's mode, with $\tau=\clmBoostMovingTauLo$--$\clmBoostMovingTauHi$ against $\clmTauLevelThree$ at rest, and fails at the same numerical endpoint ($t=\clmBoostMovingDeath$, Fig.~\ref{fig:single_throat}(d)). Time dilation ($\gamma=\clmBoostGamma$) predicts $\tau=\clmBoostDilatedTau$ in coordinate time, which the fit cannot separate from the resting value.

\subsection{The collapse endpoint}
\label{sec:single:collapse}

\begin{figure*}[t]
\centering
\includegraphics[width=0.70\textwidth]{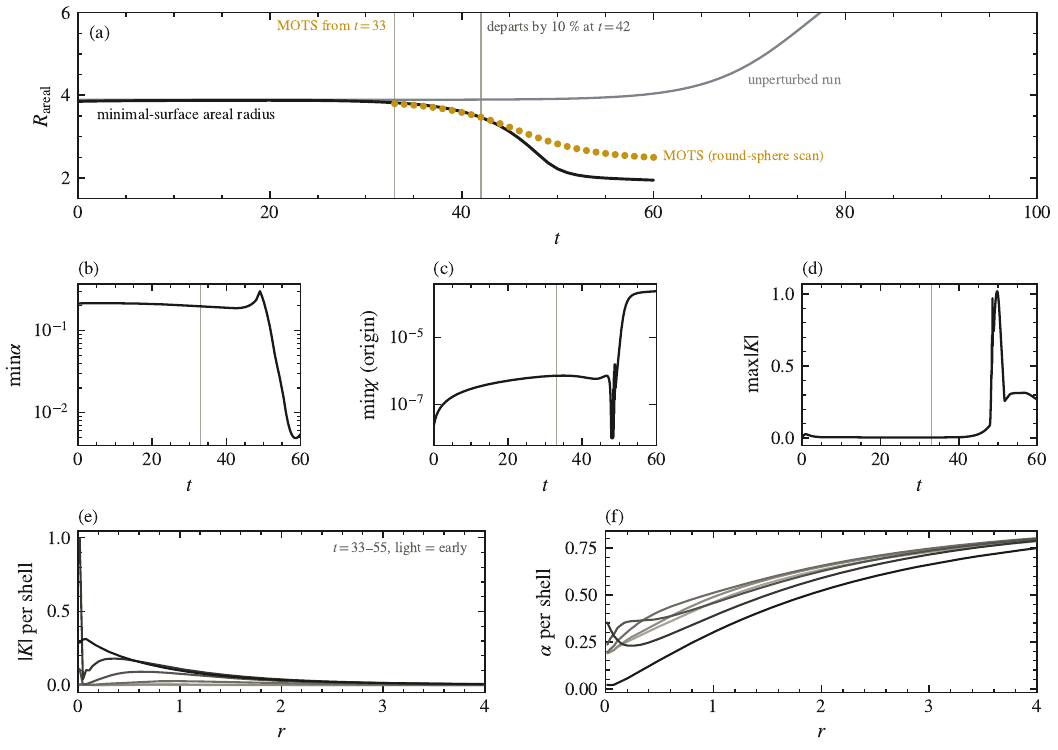}\\[4pt]
\includegraphics[width=0.70\textwidth]{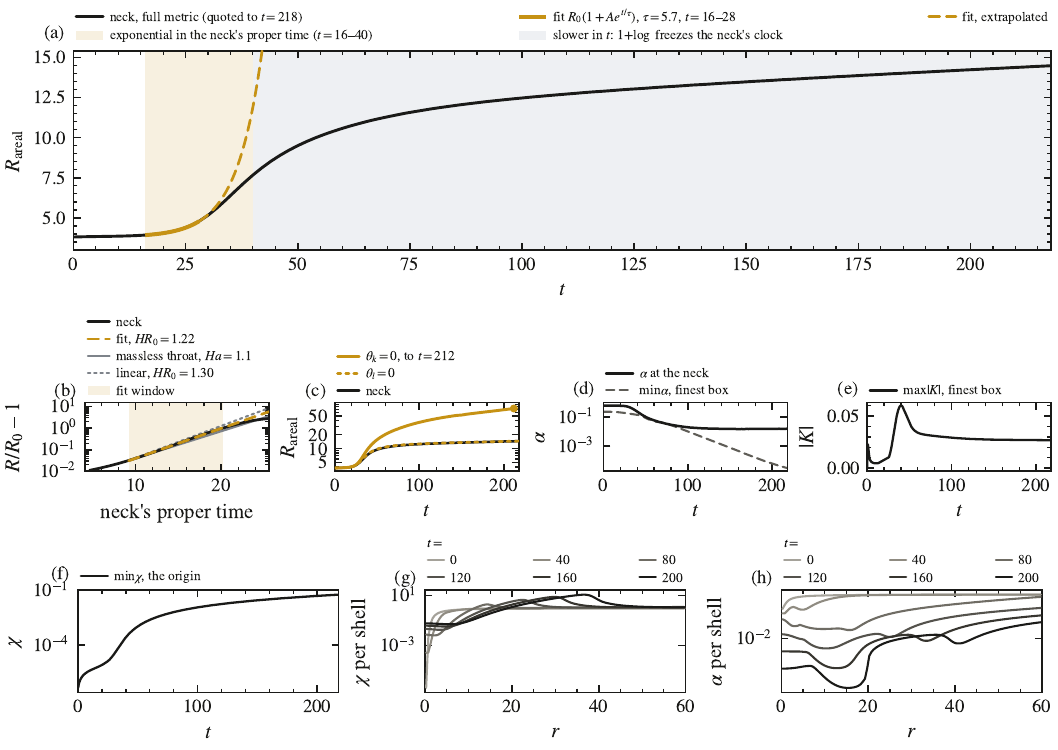}
\caption{Top: the collapsing pure-quadrupole run ($\varepsilon_2=\clmPureQuadAmpCaption$, no radial perturbation) against time. (a)~Minimal-surface areal radius (ink) and MOTS radius (gold); grey, the unperturbed level-4 run, from which this run departs by $10\%$ only at $t=\clmPureQuadDeparture$, through the quadrupole's second-order effect. (b)--(d)~Global $\min\alpha$, $\min\chi$ (at the origin, the compactified far asymptotic end) and $\max|K|$. (e),(f)~Radial profiles of $|K|$ and $\alpha$ over \clmCollapseShells{} shells, $t=33$--$55$ (light = early). Constraint norms: Fig.~\ref{fig:constraints}(c). Movie: Video~1; \href{https://youtu.be/Cc87mHcBPOA}{\nolinkurl{youtu.be/Cc87mHcBPOA}}. The throat collapses to a black hole whose horizon then shrinks. Bottom: the inflating $\varepsilon=-10^{-2}$ run against time, to $t=\clmFourWall$. (a)~Neck areal radius; gold, the exponential onset; shaded, exponential growth in the neck's proper time (gold) and the slower growth in $t$ once the $1{+}\log$ slicing freezes the neck's lapse (grey). (b)~The neck's proper-time growth rate $HR_0$, against the massless throat's~\cite{shinkai2002} and the linear mode's. (c)~Areal radii of the $\theta_-=0$ and $\theta_+=0$ surfaces. (d)~Lapse at the neck; dashed, its minimum on the finest level. (e),(f)~$\max|K|$ and $\min\chi$ on the finest level. (g),(h)~$\chi$ and $\alpha$ per shell (light = early). Constraint norms: Fig.~\ref{fig:constraints}(d). Movie: Video~2; \href{https://youtu.be/RIQCitu4tSc}{\nolinkurl{youtu.be/RIQCitu4tSc}}. The throat keeps growing and stays anti-trapped.}
\label{fig:single_collapse}\label{fig:single_inflation}
\end{figure*}

The collapse branch ends in a black hole that consumes its supporting field. On the unperturbed level-3 run the horizon's areal radius falls $\clmUnkickedHorizonRStart\to\clmUnkickedHorizonREnd$ over $\clmUnkickedHorizonSpan$ units; the $\varepsilon=+10^{-2}$ run continues without failure to $t=\clmKickedRunsClean$ behind its horizon. Ref.~\cite{shirokov2026} evolved the massless throat with the phantom's stress-energy halved; its bounce and horizon shedding are artifacts of the halving and do not occur here. The pure-quadrupole run ($\varepsilon_2=\clmPureQuadAmp$, no radial perturbation; Fig.~\ref{fig:single_collapse}) has a MOTS from $t=\clmPureQuadMotsTime$, whose radius shrinks $\clmPureQuadMotsRStart\to\clmPureQuadMotsREnd$ by $t=\clmPureQuadTrust$, while $|K|$ grows inward and the lapse collapses to zero at the centre.

The shrinking horizon is consistent with phantom accretion~\cite{babichev2004,gao2008}. In spherical symmetry a massless phantom scalar can only shrink a horizon. With the double-null metric $-2e^{-f}\mathrm{d}u\,\mathrm{d}v+R^2\mathrm{d}\Omega^2$, $T_{uv}=0$ and $T_{vv}=-(\partial_v\phi)^2$, so the Misner--Sharp mass on a marginally trapped tube obeys $\partial_v m=4\pi R^2e^{f}(\partial_v\phi)^2\partial_uR\le0$ and $\partial_u m=0$. The tube is timelike or null~\cite{hayward1996,hayward1994}, so $R=2m$ can only fall. The runs perturbed with $\varepsilon=+10^{-2}$ shrink by $\clmRegrowthShrinkPct\%$ in areal radius to a floor at $t\approx\clmRegrowthFloorTime$. Their later regrowth tracks the Hamiltonian-constraint violation of the added perturbation (Sec.~\ref{sec:model:seeds}) as it reaches the horizon; the pure-quadrupole run, whose constraint violation has no $\ell=0$ part, does not regrow (Fig.~\ref{fig:single_collapse}(a)). The shrinking takes minutes even at $10^{6}\,M_\odot$ (Table~\ref{tab:scales}). Hawking's area law assumes the null energy condition~\cite{hawking1971,ashtekar_krishnan2004}, which the phantom violates.

\subsection{The inflation branch}
\label{sec:single:fate}

The inflation branch forms no trapped surface. We evolved the $\varepsilon=-10^{-2}$ throat on an octant of an $L=\clmFourBoxL$ domain with finest $\Delta x=\clmFourFinestDx$, the resolution of our $L=\clmBoxL$ level-3 runs (Fig.~\ref{fig:single_inflation}). Its neck grows to the end of the run, by a factor $\clmFourRRatio$ at $t=\clmFourWall$, and stays anti-trapped~\cite{hayward1999}: the $\theta_-=0$ horizon sits on the neck and the $\theta_+=0$ surface moves outward, to $R\approx\clmFourHorizonR$ by $t=\clmFourHorizonTime$. Until $t\approx\clmFourSHWinHi$ the neck grows exponentially in its own proper time at a rate $H$, with $HR_0=\clmFourSHRate$ (locally $\clmFourSHLocalLo$--$\clmFourSHLocalHi$), where $R_0=\clmFourRStart$ is the perturbed neck's initial radius. This is $\clmFourLinearGap\%$ below our linear mode's $\clmFourLinearRate$, which the local rate reaches. Spherical evolutions find the same proper-time exponential: $Ha\approx\clmSHMassless$ for the massless throat~\cite{shinkai2002} and exponential growth for massive ones~\cite{gonzalez2009}. After that the $1{+}\log$ slicing freezes the lapse at the neck ($\clmFourLapseStart\to\clmFourLapseEnd$): only $\approx\clmFourTauLate$ units of its proper time pass by $t=\clmFourWall$, and its growth per unit $t$ slows. This slowdown is a property of the slicing, not of the throat.

The inflating throat sheds its own support: the monopole scalar flux through every sphere grows quasi-exponentially while the run is reliable (Sec.~\ref{sec:gw:censorship}). It draws on a store that, outside the static throat, is a Misner--Sharp deficit $R_\star/2-m=\clmInfStoreDeficit$. While the exterior slicing is still quiet, the runs of Sec.~\ref{sec:gw:censorship} shed $\clmInfShedQuiet$ of it through $R=18$, a few percent. $R_{\min}(t)$ has not turned down by $t=\clmFourWall$.

\subsection{Lifetime and physical scale}
\label{sec:single:lifetime}

A static throat survives for about $\tau\ln(1/\varepsilon)$, so its lifetime grows only logarithmically as the perturbation $\varepsilon$ shrinks. At fixed $m/a$ the solution is scale-free, and the measured e-folding time in proper time is $\tau\alpha_{\rm th}=\clmProperEfoldThree$--$\clmProperEfoldFour\,R_\star$, about one light-crossing time of the throat. The delay at equal horizon radius (Sec.~\ref{sec:single:branching}) puts the cost of each decade in $\varepsilon$ at $\clmMatchedDelayLo$--$\clmMatchedDelayHi\,M\simeq\tau\ln10$. The unperturbed throat's horizon time places the level-3 truncation error at $\varepsilon\gtrsim\clmNoiseSeed$. Table~\ref{tab:scales} converts these to physical units. No static throat of astrophysical mass survives: a stellar-mass drainhole leaves its static state within milliseconds, a supermassive one within minutes, and only at $10^9\,M_\odot$ does the lifetime reach days. The logarithm makes this robust: a perturbation ten decades smaller extends a $30\,M_\odot$ throat's lifetime only from $\clmPhysEpsTwoThirty$ to $\clmQuietLifetime$ ms.

\begin{table}[t]
\caption{Size, e-folding time $\tau=\clmTauLevelThree\,M$ (Sec.~\ref{sec:single:instability}) and time to a horizon of the fiducial throat ($R_\star=\clmRstarShort\,M$) in physical units, for four ADM masses $M$. The horizon times are measured for the two added perturbations, both outward ($\varepsilon>0$, which collapses; Fig.~\ref{fig:single_throat}(c)), and bounded from above for the unperturbed throat (Sec.~\ref{sec:single:branching}). Other amplitudes follow $t_{\rm BH}\simeq\clmHorizonTimeEpsTwo\,M+\tau\ln(10^{-2}/\varepsilon)$ at equal horizon radius: $\clmLawEpsThree\,M$ at $10^{-3}$, whose first MOTS forms smaller, at $\clmHorizonTimeEpsThree\,M$.}
\label{tab:scales}
\scriptsize
\begin{ruledtabular}
\begin{tabular}{lccccc}
 & & & \multicolumn{3}{c}{time to a horizon} \\
$M$ & $R_\star$ & $\tau$ & $\varepsilon=+10^{-2}$ & $+10^{-3}$ & unperturbed ($\le$) \\
\hline
$30\,M_\odot$ & $\clmPhysRstarThirty$ km & $\clmPhysTauThirty$ ms & $\clmPhysEpsTwoThirty$ ms & $\clmPhysEpsThreeThirty$ ms & $\clmPhysNoiseThirty$ ms \\
$10^{3}\,M_\odot$ & $\clmPhysRstarThousand$ km & $\clmPhysTauThousand$ ms & $\clmPhysEpsTwoThousand$ ms & $\clmPhysEpsThreeThousand$ s & $\clmPhysNoiseThousand$ s \\
$10^{6}\,M_\odot$ & $\clmPhysRstarMillion$ km & $\clmPhysTauMillion$ s & $\clmPhysEpsTwoMillion$ s & $\clmPhysEpsThreeMillion$ min & $\clmPhysNoiseMillion$ min \\
$10^{9}\,M_\odot$ & $\clmPhysRstarBillion$ AU & $\clmPhysTauBillion$ h & $\clmPhysEpsTwoBillion$ h & $\clmPhysEpsThreeBillion$ d & $\clmPhysNoiseBillion$ d \\
\end{tabular}
\end{ruledtabular}
\end{table}

\subsection{Type-I criticality}
\label{sec:single:critical}

In the language of critical collapse the static throat is consistent with a type-I critical solution~\cite{gundlach2007}: a static solution with a single unstable mode (Sec.~\ref{sec:single:instability}) on the threshold between two end states. Here these are collapse and inflation, where the standard problem has collapse and dispersal. Two perturbation amplitudes show both type-I signatures but do not establish the scaling. The first is time scaling: data a distance $\varepsilon$ from threshold stay near the critical solution for a time $\tau\ln(1/\varepsilon)$, which is the lifetime law of Table~\ref{tab:scales}. Both branches leave along the same mode, each $\pm\varepsilon$ pair crossing at the same $t_\times$ (Sec.~\ref{sec:single:branching}). The second is a mass gap: black-hole formation turns on at finite mass, as at the static Bartnik--McKinnon solution of Yang--Mills collapse~\cite{choptuik1996}, not at the vanishing mass of type-II scaling. The $\varepsilon=+10^{-2}$ and $+10^{-3}$ horizons form at $R=\clmHorizonRadiusEpsTwo$ and $\clmHorizonRadiusEpsThree$, both close to the throat's own $R_\star$: a tenfold smaller perturbation leaves the black hole's mass at formation almost unchanged, where type-II scaling would shrink it as a power of $\varepsilon$. Shinkai and Hayward found evidence for critical solutions with a definite black-hole mass in spherical evolutions of the massless throat~\cite{shinkai2002}; here the critical solution is the static throat itself.

\section{Two throats at rest}
\label{sec:pair}

\begin{figure*}[!t]
\centering
\includegraphics[width=0.66\textwidth]{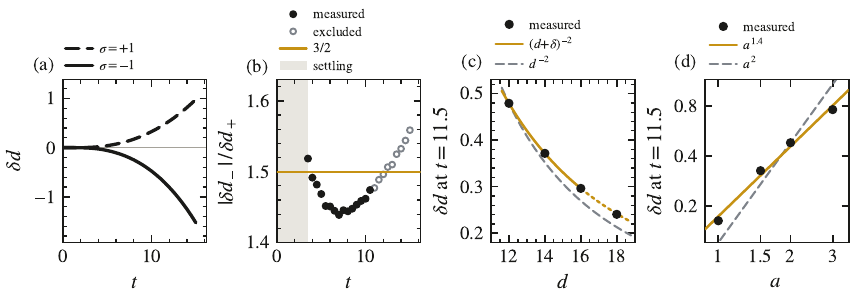}
\caption{Separation changes of throats released from rest, for matched pairs (Sec.~\ref{sec:model:solve}): the sign rule (a,b), the force law (c) and its growth with throat width (d). (a)~Separation change of two $d=\clmPairSeparation$ pairs that differ only in the sign of one scalar: like-signed ($\sigma=+1$, dashed) and opposite-signed ($\sigma=-1$, solid). Separations are inverse-$\chi$-weighted centroids of the midplane $\chi$ minima, one per throat. (b)~Pull-to-push ratio against the prediction $(Q{+}1)/(Q{-}1)=\clmScalarQRatio$ (gold), with $Q$ the scalar-to-gravity force ratio of Eq.~\eqref{eq:force}. Excluded: the gauge-settling window $t<\clmSignSettleWindow$ (grey) and $t>\clmSignWindowEnd$ (open circles), where the closing pair has closed $\clmSignWindowClosed\%$ of its separation. (c)~Separation change by $t=11.5$ of like-signed pairs at $d=12/14/16/18$ (dots), against inverse-square in $d$ scaled through $d=12$ (grey, dashed) and $A/(d+\delta)^2$ fitted on $d=12$--$16$ (gold; dotted at $d=18$). (d)~The same at $d=12$ against throat width $a=1/1.5/2/3$, on logarithmic axes, with $a^2$ scaled through $a=2$ (grey, dashed) and a power-law fit (gold). Opposite-signed throats attract, and the like-signed push falls more slowly than inverse-square in $d$ and grows more slowly than $a^2$.}
\label{fig:pair}
\end{figure*}

\subsection{The sign rule}
\label{sec:pair:sign}

Two identical throats released from rest push apart; reversing one throat's scalar ($\sigma=+1\to-1$ in Eq.~\eqref{eq:force}) turns the push into a pull (Fig.~\ref{fig:pair}(a,b)). Gravity and the scalar force share one radial law, so their ratio $Q$ does not depend on separation: no separation, mass or approach speed lets a like-signed pair fall together. The sign reversal leaves $T_{ab}$, and with it the one-body geometry, unchanged. The measured pull-to-push ratio at $a=\clmThroatA$ is $\clmMatchedSignRatio\pm\clmMatchedSignRatioErr$ (scatter over $15$ slices), against the point-charge, large-separation value $(Q+1)/(Q-1)=\clmScalarQRatio$ of Eq.~\eqref{eq:force}. The throat mass is equal in all runs, but $\lambda$ and $M_{\rm ADM}$ differ between the two runs (Table~\ref{tab:energy}), a systematic not included in this scatter.

\subsection{The force law}
\label{sec:pair:force}

To measure how the push falls with distance (Fig.~\ref{fig:pair}(c)), we released \clmCountLikeSignedPairs{} like-signed pairs from rest, identical except for the separation $d=12/14/16/18$, and compared them at a common $t=11.5$. For displacements this small, the separation change $\delta d$ at a fixed time is proportional to the force, so under the inverse-square law of Eq.~\eqref{eq:force} $\delta d\,d^2$ would be the same at every separation. Instead it rises monotonically, from $\clmFdSquaredLo$ to $\clmFdSquaredHi$, which excludes inverse-square in the coordinate separation. The law $F\propto(d+\delta)^{-2}$ describes the data: $\delta$ fitted on $d=12$--$16$ alone predicts $\clmOffsetPrediction$ at $d=18$, where inverse-square gives $\clmInverseSquarePrediction$ and the measurement is $\clmDLadderEighteen$. The fit to all four separations gives $\delta=\clmMatchedOffsetDelta$. Since $\delta$ is a sizable fraction of the mouth radius $R_\star=\clmRstarShort$, the force is inverse-square in a separation that exceeds the coordinate one by most of a throat radius. At $d=12$ the push grows with throat width $a=1/1.5/2/3$ (Fig.~\ref{fig:pair}(d)), as a power law $\delta d\propto a^n$ with $n=\clmWidthExponentLo$--$\clmWidthExponentHi$ when fitted over $t=8$--$11.5$. Point charges, whose net push $(Q-1)\,m^2/d^2$ is $a^2/d^2$ by Eq.~\eqref{eq:force}, would give $n=2$.

\subsection{The repulsion branch}
\label{sec:pair:recede}

A like-signed pair keeps receding, and each of its throats ends as an isolated throat perturbed inward does: both inflate (Video~12; \href{https://youtu.be/FnbDs-0n5zw}{\nolinkurl{youtu.be/FnbDs-0n5zw}}). Released from rest at $d=12$, the pair's separation grows from $\clmScatterSepStart$ to $\clmScatterSepSixty$ by $t=60$. Each mouth opens from $R_{\min}=\clmScatterRminStart$ to $\clmScatterRminThirty$ by $t=30$, with both null expansions positive on its minimal surface. It is therefore anti-trapped, on the inflation branch of Sec.~\ref{sec:single:fate}. The MOTS finder finds no MOTS about the mouth after $t=\clmScatterMotsLast$. The pair radiates no burst: its $\ell=2$ $\Psi_4$ keeps one sign and drifts smoothly over the whole interval in which the run is reliable.

\section{The head-on merger}
\label{sec:headon}

\subsection{Contact and the common horizon}
\label{sec:headon:contact}

\begin{figure*}[t]
\centering
\includegraphics[width=0.80\textwidth]{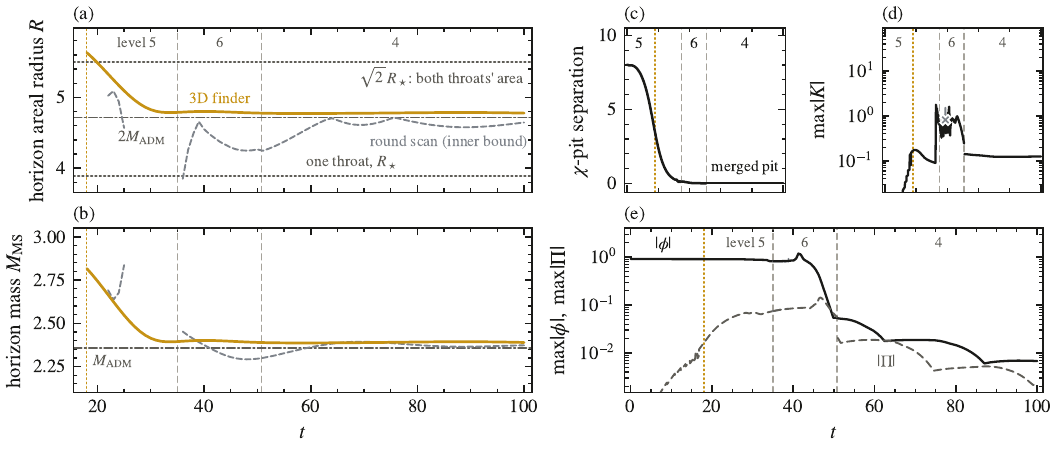}
\caption{Areal radius and Misner--Sharp mass of the common MOTS of the head-on merger against time ($d=\clmHeadonSeparation$ from rest, opposite signs). The run has three restart segments (dashed rules at $t=\clmHeadonCsmSeamOne$ and $\clmHeadonCsmSeamTwo$; levels named in each panel). Level 5 runs to a NaN in $h_{11}$ at the merged core at $t=\clmHeadonCsmNanTime$, inside the MOTS (grey after it, up to the failure cross in (c)--(e)); level 6 restarts from the $t=35$ checkpoint; level 4 continues to $t=100$ with no NaN. (a)~Areal radius and (b)~Misner--Sharp mass. Gold: the MOTS finder (Sec.~\ref{sec:setup:mots}; post-processing MOTS search, one point per output slice), with the horizon forming at $t=\clmHeadonCsmMotsTime$ with $M_{\rm MS}=\clmHeadonCsmMotsMass$ and regrowing by at most $\clmHeadonCsmRegrow\%$; at $t=100$, $M_{\rm MS}=\clmHeadonCsmEndMass$. Grey dashes: the round-sphere scan, late and small at formation and finding no surface over $t=26$--$35$ (Sec.~\ref{sec:headon:contact}). Rules: one isolated throat, $R_\star=\clmRstarShort$; $\sqrt2R_\star$, a sphere with both throats' area; $2M_{\rm ADM}$ ($M_{\rm ADM}$ in (b)). (c)~Separation of the $\chi$ minima. (d)~$\max|K|$. (e)~Grid maxima of $|\phi|$ (solid) and $|\Pi|$ (dashed). Gold dotted rule: horizon formation. Movie: Video~3; \href{https://youtu.be/3sx9GHRTYVs}{\nolinkurl{youtu.be/3sx9GHRTYVs}}. The horizon shrinks toward the Schwarzschild radius of the pair's mass.}
\label{fig:headon_collapse}
\end{figure*}

We released the pair from rest ($p=0$) at $d=\clmHeadonSeparation$ with opposite scalar signs, on matched, solved data ($M_{\rm ADM}=\clmHeadonCsmAdmMass$). The pair falls together, and the MOTS finder (Sec.~\ref{sec:setup:mots}) finds a common MOTS around both mouths from $t=\clmHeadonCsmMotsTime$ (Fig.~\ref{fig:headon_collapse}). It forms while the $\chi$ minima are still ${\sim}\clmHeadonCsmBirthSep$ apart, with $R=\clmHeadonCsmMotsRadius$ and $\clmHeadonCsmBornArea$ times the two throats' summed area. No surface ever closes about either mouth alone: the horizon forms around both mouths at once. At formation $R/2=M_{\rm MS}$ exceeds $M_{\rm ADM}$ by $\clmHeadonCsmPenrose\%$, violating the Penrose inequality. The phantom permits this, since the inequality presupposes the non-negative energy density it lacks~\cite{huisken2001,mars2009}. Each throat's own Misner--Sharp mass, $R_\star/2=\clmThroatOwnMass$, already exceeds its ADM mass by the negative energy of the field outside it. After formation the horizon shrinks overall (Sec.~\ref{sec:headon:wall}). The round-sphere scan finds the horizon $\clmHeadonScanLate$ units late and $\clmHeadonScanSmall\%$ small, and it finds no surface while the horizon is strongly deformed (Fig.~\ref{fig:headon_collapse}(a)); this is the scan-versus-MOTS offset of Sec.~\ref{sec:setup:diagnostics}.

\subsection{The remnant black hole}
\label{sec:headon:wall}

The remnant's horizon shrinks as it swallows the phantom field that held the throats open (Fig.~\ref{fig:headon_collapse}(e)). It loses $\clmHeadonCsmShrinkArea\%$ of its initial area, falling to $R=\clmHeadonCsmEndRadius$ at $t=100$. Gravitational waves carry only $\clmHeadonGwShare\%$ of the Misner--Sharp mass it loses: the radiated gravitational-wave energy is $E_{\rm GW}=\clmDetEHeadon\,M$ (Sec.~\ref{sec:headon:wave}). The rest is the negative-energy field falling into the black hole. The level-5 run fails (a NaN) inside the horizon at $t=\clmHeadonCsmNanTime$, the interior failure of Sec.~\ref{sec:spiral:wall}. Restart segments at level 6 and then level 4 carry it to $t=100$, and the horizon's radius is continuous across the restart times.

The first law of Sec.~\ref{sec:single:collapse}, evaluated on this surface, shrinks it under a phantom influx ($T_{ab}k^ak^b<0$) and grows it through the outgoing shear. Over $t=51$--$60$, rerun with every output slice kept, the law brackets the observed $\Delta R=\clmHeadonFirstLawMeasured$: the phantom influx alone gives $\clmHeadonFirstLawPhantom$, and adding the shear's growth term gives $\clmHeadonFirstLawShear$. By $t=90$--$100$ the influx is spent: the phantom rate on the surface is $\clmHeadonCsmFluxFold$-fold below its $t=45$--$55$ value, while the radius still falls at $\clmHeadonCsmSlope$ per unit time.

At $t=100$ the horizon's radius is $\clmHeadonCsmEndOverSchw\%$ above $2M_{\rm ADM}$, and it is still shrinking. If it settles at the pair's Bondi mass, then $R=2(M_{\rm ADM}-E_{\rm GW}-E_\phi)$, where $E_\phi$ is the radiated scalar-field energy. With the measured $E_{\rm GW}$ this gives $E_\phi\approx-\clmHeadonCsmScalarImplied$, of the sign and order of the flux integrals of Sec.~\ref{sec:gw:censorship}. In spherical symmetry, a collapsed Ellis--Bronnikov throat settles to Schwarzschild on both sides once it has expelled the phantom field~\cite{kain2026}. The $d=6$ merger's remnant lies within $\clmMergerRemnantAgree\%$ of the same radius at $t=100$ (Sec.~\ref{sec:spiral:inspiral}), consistent with an end state set by the pair's mass.

\section{The orbital channel}
\label{sec:spiral}

\subsection{Merger or fly-by}
\label{sec:spiral:capture}

We varied the tangential momentum $p$ at $d=12$ on boosted, solved data (Videos~4--6, at $p=\clmSpiralMomentumTwentyFive$, $\clmEgwPSixty$ and $\clmEgwPNinety$: \href{https://youtu.be/fh3pU7xCU_I}{\nolinkurl{youtu.be/fh3pU7xCU_I}}, \href{https://youtu.be/-ejFKpetL1Y}{\nolinkurl{youtu.be/-ejFKpetL1Y}}, \href{https://youtu.be/yzvwjC46lJY}{\nolinkurl{youtu.be/yzvwjC46lJY}}). At $p=\clmSpiralMomentumTwentyFive$ the pair passes at closest separation $\clmFlybyLbfClosest$ and recedes. At $p=\clmSpiralMomentumFlyby$ it passes closer, at $\clmPFortyFiveClosest$, and also recedes. Both fly-bys reach their stop times with no NaN, and the MOTS finder finds no trapped surface on either. The slowest run, $p=\clmSpiralMomentumProd$, reaches contact (Sec.~\ref{sec:spiral:race}). The two fastest plunge to a merged core on which no common MOTS converges (Sec.~\ref{sec:spiral:wall}). The plunge boundary thus lies between $\clmPFortyFiveCircFrac\%$ and $\clmPSixtyCircFrac\%$ of the circular momentum under the combined pull of Eq.~\eqref{eq:force}, $1+Q$ times gravity's, and so brackets the circular value. This pull is central, so a pair that misses contact keeps its angular momentum and swings past. From $p=\clmSpiralMomentumTwentyFive$ up, closest approach shrinks as $p$ grows, whereas a Newtonian pericentre grows with $J=pd$. Capture here is therefore contact, not an innermost orbit. Both fly-by pairs start below the circular momentum and stay bound in Newtonian terms. What follows a pass is set by the throats, which inflate (Sec.~\ref{sec:spiral:race}).

The recession after the turn is not that of two compact throats. On both fly-bys the mouths inflate, and the runs stop being reliable as they do, at $t=\clmFlybyLbfTrust$ and $\clmPFortyFiveTrust$ (Sec.~\ref{sec:setup:diagnostics}); nothing later is quoted, from the runs or their waves (Sec.~\ref{sec:gw:flyby}).

\subsection{Merger against throat inflation}
\label{sec:spiral:race}

No $d=12$ pair forms a common horizon. The slowest run, $p=\clmSpiralMomentumProd$, reaches contact at $t\approx\clmPTwelveContactTime$ and inflates. The MOTS finder, run from $t=\clmPTwelveMotsFrom$ until the run stops being reliable at $t=\clmPTwelveTrust$, finds no common MOTS. In a test run with inward momentum ($p=\clmDTwelveInwardP$) the mouths inflate faster than the orbit closes. Their $R_{\min}$ grows by $\clmDTwelveInflTwentyFour\%$ by $t=24$, while the infall stalls near separation $\clmDTwelveStallSep$, and the finder finds no common MOTS on any output slice.

The companion selects the branch, as an added perturbation does for an isolated throat (Sec.~\ref{sec:single:branching}), which collapses when perturbed outward and inflates when perturbed inward. The companion acts on each mouth like an inward perturbation. Each mouth starts in the isolated throat's shape (Table~\ref{tab:placement}) yet grows like an isolated throat with $\varepsilon_{\rm eff}\approx-\clmRaceKickTwelve\%$ for $d=12$ at rest and $-\clmRaceKickEight\%$ for $d=8$. Every pair's mouths therefore start on the inflation branch, at any momentum, and grow on the throat's e-folding time $\tau\approx\clmTauLevelThree$ (Sec.~\ref{sec:single:instability}). A merger is a race against this inflation. The common horizon forms after $\clmRaceFoldsHeadon$ e-folding times for the head-on and $\clmRaceFoldsSix$ for the $d=6$ merger (Sec.~\ref{sec:spiral:inspiral}). The $d=12$ approach takes $\clmRaceFoldsTwelve$ e-folding times to its closest pass and forms no horizon. An off-axis merger therefore needs a short separation and a small tangential momentum.

\subsection{The $d=6$ merger}
\label{sec:spiral:inspiral}

\begin{figure*}[t]
\centering
\includegraphics[width=0.80\textwidth]{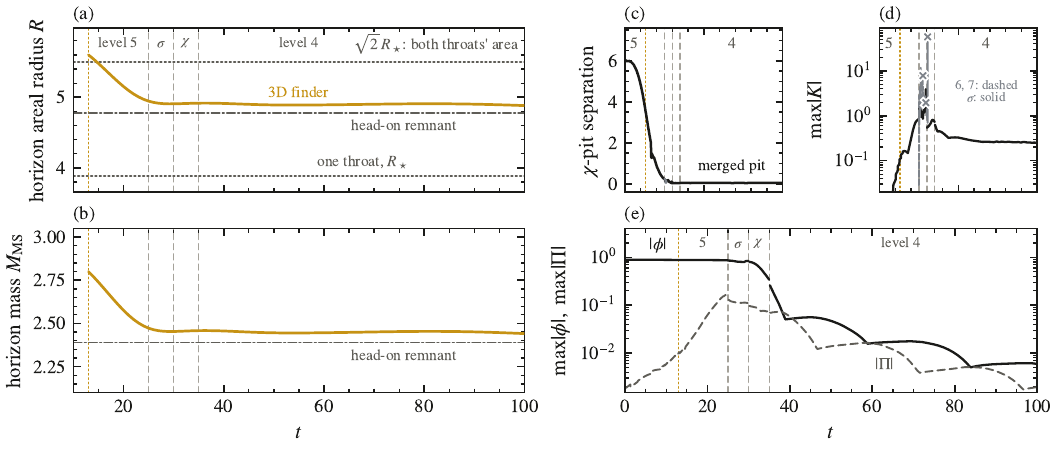}
\caption{Common-horizon properties, separation, $\max|K|$ and field maxima against time for the off-axis merger ($d=6$, tangential $p=\clmDSixP$, opposite signs, boosted, constraint-solved data). The run has four restart segments (faint dashed rules at the restart times $t=\clmMergerSeamOne$, $\clmMergerSeamTwo$ and $\clmMergerSeamThree$): level 5, $\sigma_{\rm KO}=1.0$, the $\chi$ floor, and level 4 to $t=100$. (a)~Areal radius and (b)~Misner--Sharp mass of the common MOTS (gold; Sec.~\ref{sec:setup:mots}); dash-dotted rules, the head-on remnant's values. Dotted rules in (a): $R_\star$ of one isolated throat and $\sqrt2R_\star$, the radius of both throats' total area. (c)~Separation of the $\chi$ minima (grey: segment 1 past its restart time). (d)~$\max|K|$ of every run: the segments used in ink, the level-6 and level-7 restarts grey dashed, the $\sigma_{\rm KO}=1.0$ restart grey solid; crosses mark failures. (e)~Grid maxima of $|\phi|$ (solid) and $|\Pi|$ (dashed). Gold dotted rule: horizon formation. Movie: Video~7, \href{https://youtu.be/jNnMAXPjyRA}{\nolinkurl{youtu.be/jNnMAXPjyRA}}. The common horizon shrinks toward the head-on remnant's radius and mass, and it swallows the field.}
\label{fig:spiral_collapse}
\end{figure*}

We therefore chose $d=6$ with a small tangential momentum, $p=\clmDSixP$ per mouth (about $\clmDSixCircFrac\%$ of circular). A level-3 test run merges, and the main run repeats it in the $L=128$ domain (Fig.~\ref{fig:spiral_collapse}). The MOTS finder finds a common MOTS around both mouths from $t=\clmMergerMotsTime$, while the $\chi$ minima are still $\clmMergerBirthSep$ apart. It forms with $R=\clmMergerMotsRadius$ and shrinks by $\clmMergerShrinkR\%$ to $R=\clmMergerEndRadius$ at $t=100$, still falling. The result is a black hole reached from an off-axis approach. As in the head-on, the horizon forms around both mouths at once and swallows the pair mid-inflation.

The run was restarted three times to pass the interior failure (Sec.~\ref{sec:spiral:wall}): level 5 from $t=0$ to $t=\clmMergerLegOneStop$; $\sigma_{\rm KO}=1.0$ to $t=\clmMergerSigDeath$; a $\chi$ floor ($\min\chi=10^{-4}$) to $t=\clmMergerChiStop$; and level 4 with both settings reverted to $t=100$, with no NaN. The floor acts only inside the horizon, over $t=\clmMergerSeamTwo$--$\clmMergerChiStop$. Every quoted number comes from the finder's surface outside it.

\subsection{Interior failure after horizon formation}
\label{sec:spiral:wall}

Two failure modes appear in merged cores, and each is removed by a change of setting. The first is a NaN in the metric component $h_{11}$ at the core's centre, with the global constraint norms clean to the last step. It occurs on the level-5 head-on run, on the $d=6$ run and on the $p=\clmEgwPSixty$ plunge. Refinement does not remove it: on the $d=6$ run the level-7 restart fails earlier than the level-6 one ($t=\clmMergerLvlSevenDeath$ against $\clmMergerLvlSixDeath$), because a finer grid follows the collapse further. Raising the dissipation to $\sigma_{\rm KO}=1.0$ does not remove it either. What removes it is the $\chi$ floor: with $\min\chi$ capped at $10^{-4}$, the runaway of the $1/\chi$ terms cannot complete. With the floor, both the $d=6$ run and the $p=\clmEgwPSixty$ plunge pass their failure times and reach $t=100$. The second mode is a runaway of $\max|K|$ at the merging core, from $\clmPNinetyKLo$ to $\clmPNinetyKHi$ in one step on the $p=\clmEgwPNinety$ run at $t=\clmPNinetyDeath$. Level 6 removes it on the $d=6$ run. On the $p=\clmEgwPNinety$ run a level-5 restart removes it, but that run fails at $t=\clmPNinetyLvlFiveDeath$.

Where a horizon forms, the failure occurs inside it. The core fails only after a common MOTS encloses it: the $d=6$ NaNs arrive $\clmMergerWallLeadLo$--$\clmMergerWallLeadHi$ units after its horizon forms, and the head-on's $\clmHeadonCsmWallLead$ after its. Isolated throats that collapse at level 3 also fail behind their horizons ($\varepsilon=+10^{-3}$, $\clmKickThreeLead$ units after; $\varepsilon=+10^{-2}$ combined with $\varepsilon_2=10^{-2}$, $\clmDressedLeadLevelThree$ after), and none fails at level 4. The failure is local, the exterior is already set, and the finder's surface passes every failure time unmoved. What diverges is the coordinate representation of a merged interior, behind the horizon.

Our simulations thus test weak cosmic censorship outside the physically reasonable matter it is posed for~\cite{penrose1969,wald1997}: every resolved collapse fails, if at all, only inside a MOTS. This is evidence for censorship under null-energy violation, not a proof, since the failure is a coordinate blow-up on the slice and not a located curvature singularity. The plunges whose horizon never converges (Sec.~\ref{sec:spiral:capture}) leave their failures outside any horizon on the slice, at the edge of what the grid resolves.

\section{Gravitational waves}
\label{sec:gw}

\begin{figure*}[t]
\centering
\includegraphics[width=0.80\textwidth]{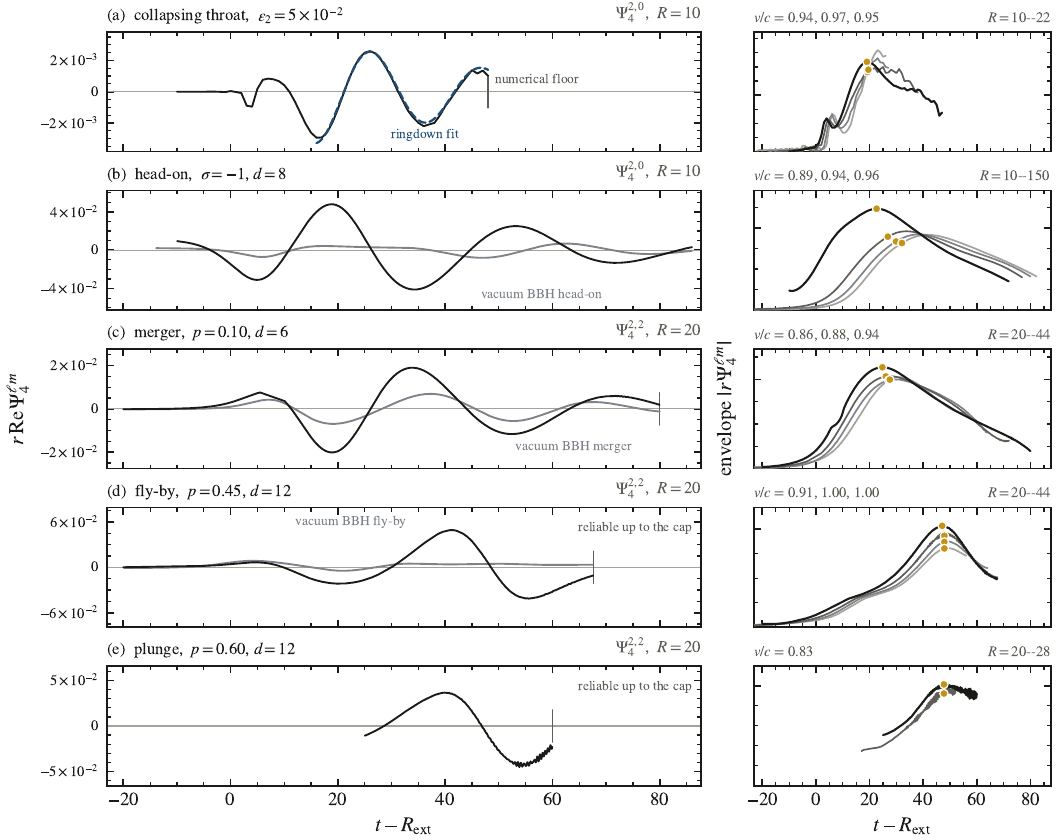}
\caption{Gravitational waveforms of every scenario against retarded time, one row per run. Left: $r\,\mathrm{Re}\,\Psi_4$ of the dominant mode at the innermost sphere, each row on its own scale. Row (a) is the collapsing throat at $\varepsilon_2=\clmGwSeedStrong$, with the ringdown fit of Sec.~\ref{sec:single:gw} dashed. Caps mark where each drawn waveform ends: for the collapsing throat, before its numerical floor exceeds $\clmGwDrawFloorFrac\%$ of the burst peak; for the fly-by and the plunge, where each run stops being reliable. Row (b) is the head-on run in the larger domain ($L=\clmFarzoneBoxL$, spheres to $R=\clmFarzoneOuterR$). The merger's waveform runs through the ringdown, interpolated linearly across one restart time. Grey under (b)--(d): the vacuum controls, black holes at the same $d$ and $p$, on the same scale. Right: the amplitude envelope at every sphere against its own retarded time (ink to faint, inner to outer), the upper envelope of $|r\Psi_4|$ for the complex $(2,2)$ modes and the analytic-signal envelope for the real $(2,0)$ modes. Gold dots mark the correlated wavefront (Sec.~\ref{sec:controls:bbh}). Movies of the head-on run in the larger domain and of the vacuum head-on and fly-by: Videos~11 \href{https://youtu.be/vesOxaP5GSg}{\nolinkurl{youtu.be/vesOxaP5GSg}}, 8 \href{https://youtu.be/aRWvD4W8OOg}{\nolinkurl{youtu.be/aRWvD4W8OOg}} and~9 \href{https://youtu.be/65_3kw4obvE}{\nolinkurl{youtu.be/65_3kw4obvE}}. Every signal is outgoing and reaches the spheres in order of radius.}
\label{fig:gw_gallery}
\end{figure*}

\begin{figure*}[t]
\centering
\includegraphics[width=0.80\textwidth]{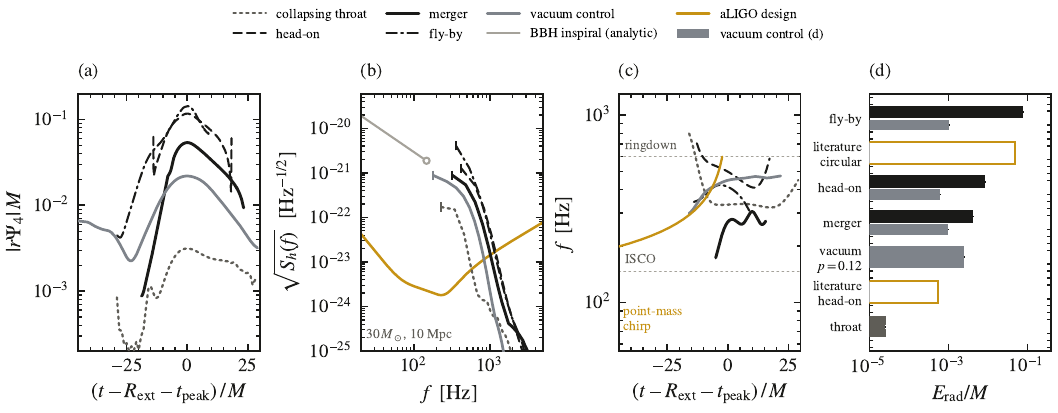}
\caption{Waveforms, spectra, frequencies and energies of every run, for a source of total mass $\clmGwLigoMass\,M_\odot$ at $\clmGwLigoDist$ Mpc, each in units of its own total mass. The vacuum control is two black holes with the separation and momentum of the $d=12$, $p=\clmSpiralMomentumProd$ run (Sec.~\ref{sec:spiral:capture}). (a)~Upper envelopes of $|r\Psi_4|M$ at the innermost sphere, aligned at each peak. (b)~Strain amplitude spectral densities against the Advanced LIGO (aLIGO) design sensitivity (gold), each from its own low-frequency corner (one cycle per waveform, ticked). The curves are falling power laws with no resolved maximum, so any apparent peak is the corner frequency $1/T$ of a waveform of duration $T$. Faint line: the Newtonian inspiral of the same binary, $|\tilde h|\propto f^{-7/6}$, up to its innermost stable circular orbit (ISCO) at $\clmGwIscoHz$ Hz. (c)~Instantaneous frequency against the point-mass chirp (gold), the ISCO and the equal-mass Kerr ringdown ($\clmGwKerrHz$ Hz); the vacuum control climbs the chirp and peaks at $fM=\clmGwTwinPeakFM$, with $f$ the frequency and $M$ the total mass (Sec.~\ref{sec:ligo:ringdown}). (d)~Energy radiated in the dominant multipole at the outermost sphere covering each waveform's window, the tick its change from the next sphere in; grey under each source, its vacuum control at the same $d$ and $p$. Open gold bars: published equal-mass vacuum values, all multipoles, for head-on infall from rest at infinity, $\clmGwLitHeadonInfall$~\cite{witek2010}, and for quasi-circular inspiral and merger, $\clmGwLitCircular$~\cite{scheel2009}. Each energy integrates $|\int r\Psi_4\,\mathrm{d}t'|^2$ over one retarded window, ending at the end of its reliable interval or where noise dominates, and is doubled for $\pm m$; gravitational channel only. Movie of the vacuum control: Video~10, \href{https://youtu.be/NkcpGiJIxXI}{\nolinkurl{youtu.be/NkcpGiJIxXI}}. Only the vacuum control climbs the chirp; no wormhole run does.}
\label{fig:gw_ligo}
\end{figure*}

Every channel radiates. Of the energies measured, the horizonless fly-by's at the capture boundary is the largest, and the plunge's is a lower bound. Figure~\ref{fig:gw_gallery} shows the peak $|r\Psi_4|$ of each channel. Each signal propagates outward at $v/c=\clmGwSpeedLow$--$\clmGwSpeedHigh$ between neighbouring spheres (Fig.~\ref{fig:gw_gallery}, right; Sec.~\ref{sec:controls:bbh}). The inner spheres lie in the near zone: the dominant periods are comparable to the extraction radii, so the peak $r\Psi_4$ falls across the spheres. The radiated energy converges outward instead (Fig.~\ref{fig:convergence}(d) in Appendix~\ref{app:health}). On the head-on run it is constant to $\clmNearZoneHeadonOuterSpread\%$ on the outer spheres and exceeds that by $\clmNearZoneHeadonInner\%$ only at the innermost. Each energy is therefore read at the outermost sphere that covers its window, with its change from the next sphere in as its error (Fig.~\ref{fig:gw_ligo}(d)). The waveforms drawn in Sec.~\ref{sec:ligo} are the innermost sphere's, whose near-zone excess in rms amplitude is $\clmNearZoneAmpExcess\%$.

Each burst arrives with its physical trigger (collapse, contact or closest pass), tens of units after the first light-crossing time $t\approx R$ (Fig.~\ref{fig:gw_gallery}, right). The transient of the constraint-solved initial data is small: on every binary the first light-crossing window, $t-R\le10$, carries at most $\clmJunkShareMax\%$ of the quoted band energy. For the perturbed single throats the wave comes from the throat, not from the initial shell of the perturbation (Sec.~\ref{sec:single:gw}).

Each channel is compared with black holes at its own $d$ and $p$ (grey in Fig.~\ref{fig:gw_gallery}(b)--(d)). The vacuum merger peaks at the drainhole merger's retarded time, $\clmGwVacMergerPeakRatio\times$ lower. The fly-by's control is unbound in vacuum: the black holes start at periapsis and coast apart, radiating a single cycle. The drainhole pair instead falls in to $d=\clmPFortyFiveClosest$ and outradiates it by $\clmGwVacPassRatioLo$--$\clmGwVacPassRatioHi\times$. For scale, the closest non-merging black-hole encounters computed in vacuum radiate ${\approx}\clmGwLitScatter$ of their total mass in all multipoles~\cite{damour2014}.

\subsection{Collapsing throat}
\label{sec:single:gw}

An isolated throat radiates only if its collapse is not spherical. The spherically perturbed run emits nothing above the $\ell=2$ numerical floor. Adding a quadrupole of amplitude $\varepsilon_2=\clmGwSeedStrong$ gives a burst of peak $|r\Psi_4|=\clmGwPeakThroat$, $\clmGwThroatOverControl\times$ the floor of the level-4 control, carrying $\clmGwEnergyThroat\,M$ (integrated to $t=\clmGwThroatEnergyCut$, before the floor dominates). The burst moves outward at $\clmGwThroatSpeedLo$--$\clmGwThroatSpeedHi\,c$ and is linear in the perturbation to $\clmGwLinDevLo$--$\clmGwLinDevHi\%$ over $\varepsilon_2=\clmGwSeedWeak$--$\clmGwSeedStrong$. With the radial perturbation removed, the quadrupole radiates the same amplitude to $\clmGwPureVsKicked\%$, only earlier: the deformation sets the wave and the radial perturbation only its timing. We read the ringdown at $\varepsilon_2=\clmGwSeedStrong$, since the weaker perturbation's burst stands only a few times above the numerical floor. The remnant rings at period $\clmGwThroatPeriod\,M$ (dashed in Fig.~\ref{fig:gw_gallery}(a)), with no drift across the window. This is the Schwarzschild fundamental mode of a black hole of mass $\clmGwThroatRingMass\pm\clmGwThroatRingMassErr$~\cite{berti2009}, near the horizon's late mass ($\clmGwThroatMassFloor$--$\clmGwThroatMassEnd$) rather than its mass at the window's start ($\clmGwThroatMassStart$).

\subsection{Head-on ringdown}
\label{sec:headon:wave}

The head-on signal is a ringing with no chirp. The $(2,0)$ mode rings with a half-period stretching from $\clmGwHeadonHalfLo$ to $\clmGwHeadonHalfHi$ and moves outward at $\clmGwHeadonSpeed\,c$ across all three spheres. Its frequency falls with time (Fig.~\ref{fig:gw_ligo}(c)). Band-integrated about its peak, it carries $E_{\rm GW}/M=\clmHeadonCsmGwEnergy$ at the outermost sphere. The vacuum control, two equal black holes released from rest at the same separation, rings in the same $(2,0)$ mode with $\clmBbhHeadonGwRatio\times$ less energy and crests $\clmBbhHeadonPeakLag$ units later. Figure~\ref{fig:gw_ligo}(d) compares it with the published value for infall from rest at infinity~\cite{witek2010}. Without the scalar's attraction, the holes fall more slowly and collide less violently.

\subsection{Merger burst}
\label{sec:spiral:burst}

The $d=6$ merger's $(2,2)$ burst is launched at contact, when the common MOTS forms at $t=\clmMergerMotsTime$ (Sec.~\ref{sec:spiral:inspiral}). It peaks at $|r\Psi_4|=\clmGwPeakSpiral$ on the innermost sphere, and its envelopes fold onto one curve across the spheres. The remnant's ringdown carries the waveform to $t=100$ on the final restart segment (Fig.~\ref{fig:gw_gallery}(c)).

\subsection{Fly-by and plunge bursts}
\label{sec:gw:flyby}

Each orbital run emits one $(2,2)$ burst at its closest pass or plunge; the fly-bys form no remnant and show no ringdown. Through the outermost sphere that covers each burst ($R=\clmExtractRadiusTwentyEight$; for the plunge, $R=\clmExtractRadiusTwenty$), $E_{\rm GW}/M=\clmEgwEnergyTwoFive$ at $p=\clmSpiralMomentumTwentyFive$, $\clmEgwEnergyFortyFive$ at $p=\clmSpiralMomentumFlyby$ (damped run, Sec.~\ref{sec:setup:grid}), and at least $\clmEgwEnergySixty$ at $p=\clmEgwPSixty$. Each is in units of its own pair's measured mass (Sec.~\ref{sec:model:boost}). At the capture boundary (Sec.~\ref{sec:spiral:capture}) the pass is deepest and slowest: the $p=\clmSpiralMomentumFlyby$ pulse carries $\clmEgwBoundaryOverScatter\times$ the shallower fly-by's energy and $\clmEgwBoundaryOverPlunge\times$ the plunge's lower bound. Each fly-by's pulse is integrated to the end of its run's reliable interval ($t-R\le\clmEgwTrustTwoFive$ and $\clmGwDrawFlybyRetarded$). The plunge run is reliable only to $\clmEgwGateSixty$ units past its $R=20$ peak, so its energy is a lower bound and the boundary-to-plunge ratio an upper bound.

\subsection{Vacuum control}
\label{sec:controls:bbh}

The chirping reference is a vacuum binary of equal masses at $d=\clmPairSeparation$ and $p=\clmSpiralMomentumProd$, with no scalar field. It merges, climbs the point-mass chirp from $\clmGwTwinFreqLo$ to $\clmGwTwinFreqHi$ Hz and turns toward the equal-mass Kerr mode (Sec.~\ref{sec:ligo:ringdown}). It also calibrates the wavefront timing: its flat-topped merger envelope defeats peak timing, which reads $\clmGwSpeedTwinPeak{}c$ on it against $\clmGwSpeedTwin{}c$ from its whole-waveform lag. Every speed quoted here except the plunge's is therefore a whole-waveform lag.

\subsection{Scalar radiation and its sign}
\label{sec:gw:scalar}

The Weyl scalar $\Psi_4$ does not measure scalar radiation. Two opposite scalar charges (every merging pair here, Sec.~\ref{sec:pair}) carry a dipole, whereas a vacuum binary's leading moment is the quadrupole. On every binary the scalar field is extracted on the same spheres as the post-processing $\Psi_4$ (Sec.~\ref{sec:setup:diagnostics}), so the two channels compare sphere by sphere (Fig.~\ref{fig:scalar_channel} in Appendix~\ref{app:supp}). The recorded kinematic flux $F_{\rm kin}$ approximates the scalar energy crossing $S_R$ per coordinate time, the flux through the world tube of $S_R$, of the canonical energy current $J^\mu\equiv-T^{(\phi)\mu}{}_{\nu}\,t^\nu$, with $t^\mu=\alpha n^\mu+\beta^\mu$~\cite{clough2021,gourgoulhon2012}:
\begin{equation}
\begin{split}
F^{\rm geo}_\phi&=\oint_{S_R}\alpha\,s_iJ^i\sqrt{q}\;\mathrm{d}^2x\\
&=-\oint_{S_R}\alpha\,\partial_t\phi\;s^j\partial_j\phi\,\sqrt{q}\;\mathrm{d}^2x\;+\;O(\beta),
\end{split}
\label{eq:scalarflux}
\end{equation}
positive for an outgoing canonical wave ($s^i$ is the sphere's outward unit normal, $\sqrt{q}$ its area element, and $\partial_t\phi=\alpha\Pi+\beta^i\partial_i\phi$). At $R=\clmGwScalarRadiusOuter$ the background's area mean of $\alpha^2\chi^{-1/2}$ is $\approx\clmGwScalarFluxFactor$, so $F_{\rm kin}$ overstates the geometric flux by about $\clmGwScalarFluxCorr\%$, neglecting shift terms.

The channel is dipole-dominated and comparable to the gravitational one. On the fly-by at $R=\clmGwScalarRadiusOuter$, $\ell=1$ carries all but one part in $10^{\clmGwScalarDipoleShare}$ of the scalar power, since symmetry forbids $\ell=0$ and $\ell=2$ for $\sigma=-1$; the remainder measures numerical symmetry breaking. It leaves as a single outward pulse, complete within the sphere's reliable interval (Fig.~\ref{fig:scalar_channel}(a)). Over that interval $|E_\phi|=\clmGwScalarEnergyFlyby\,M$, $\clmGwScalarRatioFull$ times the gravitational energy that crosses the sphere with it (Fig.~\ref{fig:scalar_channel}(a)). Part of that energy is not radiation. Before the pulse, $F_{\rm kin}$ rises on every sphere at once, ahead of any signal from the throats, and falls with radius as $R^{-\clmScalarNearSlope}$. This is the near field of the moving charges, $\clmScalarPrearchShare\%$ of $|E_\phi|$; the pulse alone gives $|E_\phi|/E_{\rm GW}=\clmScalarArchRatio$. Below we use only the channel's multipole, direction, sign and order of magnitude.

Gravity couples to minus the canonical stress tensor, Eq.~\eqref{eq:phantom}, so the physical energy flux is $-F_{\rm kin}$. The outgoing $\ell=1$ wave therefore carries negative energy, on the fly-by and, once their horizons form, on the head-on and the $d=6$ merger. If the ADM energy balance holds with this term, a bound pair emitting it gains energy and radiation reaction is anti-damped. This is the classical form of the standard objection to ghosts~\cite{carroll2003,cline2004}. The head-on remnant, slightly above $M_{\rm ADM}$ and still shrinking at $t=100$, is consistent with this balance (Sec.~\ref{sec:headon:wall}). The dipole's physical angular-momentum flux is opposite to that of the gravitational waves, so it returns angular momentum to the orbit. It does not invert the nonlinear memory~\cite{christodoulou1991,thorne1992,favata2009}. Per unit energy, an equatorial dipole projects onto the $(2,0)$ memory with the opposite sign to a quadrupole, so the negative scalar flux adds to the gravitational memory, by $\clmMemoryRatioLo$--$\clmMemoryRatioHi$ of it on the fly-by.

\subsection{Scalar emission after horizon formation}
\label{sec:gw:censorship}

In the head-on run the scalar-field extraction covers the whole evolution, and the horizon switches the channel off (Fig.~\ref{fig:scalar_censorship} in Appendix~\ref{app:supp}). After the common MOTS forms ($t=\clmHeadonCsmMotsTime$), the mouths' remaining scalar charge leaves as a single $\ell=1$ pulse along the collision axis. The pulse crests later at each sphere in causal order, and its envelope then decays. Underneath, the remnant's scalar dipole rings down. Fitted from $t=40$ as one damped oscillation, its frequency agrees to $\clmScalarQnmFreqDev\%$ with the $\ell=1$ scalar quasinormal mode of a Schwarzschild black hole of the final MOTS mass~\cite{berti2009}, for fits started anywhere over $t=35$--$45$. Its damping is $\clmScalarQnmDampDev\%$ slower. Integrated from where each sphere's flux turns outgoing to $t=\clmGwHeadonNoiseGate$, the energies keep the sign of Sec.~\ref{sec:gw:scalar}.

The $d=6$ merger, the only other encounter that forms a horizon (common MOTS from $t=\clmMergerMotsTime$; Sec.~\ref{sec:spiral}), loses its scalar charge the same way over its restart segments. Its dipole crests and then falls exponentially, with e-folding time $\clmGwCensMergerTau$. Without a horizon nothing switches the channel off. The fly-by radiates its pulse as the mouths swing past one another, and the emission ends with the encounter itself. The mouths leave intact, still carrying the charges that source the dipole.

The isolated throat shows the same contrast. We evolved the $\varepsilon=\clmGwKickedEps$ throat again at level \clmGwKickedLevel{} with scalar-field extraction. It collapses at the expected time, with a permanent MOTS from $t=\clmGwKickedMotsTime$, and its scalar envelope falls across the spheres after the burst, with late e-folding times of $\clmGwKickedEfoldLo$--$\clmGwKickedEfoldHi$. The unperturbed level-\clmGwSingleLevel{} throat inflates with no trapped surface found by the round-sphere scan or the MOTS finder. Its $\ell=0$ scalar monopole instead grows quasi-exponentially, with e-folding time ${\approx}\clmGwSingleEfold$. The same run with the domain doubled to $L=\clmGwSingleBox$ agrees to $\clmGwSingleAgreePct\%$, so the growth is the throat's and not the boundary's or the sponge's. The constraint norms grow later, so we use only $t\lesssim\clmGwSingleCountTime$. The contrast cannot be separated from the end state: the runs without a horizon are also those whose mouths inflate toward the spheres. What we measure is that the scalar channel shuts off behind a horizon and grows while a throat inflates.

\section{Observational signatures}
\label{sec:ligo}

\subsection{The frequency track}
\label{sec:ligo:ringdown}

The frequency track separates a drainhole burst from a binary black hole by what comes before the ringdown (Fig.~\ref{fig:gw_ligo}(c)). Waveform shape alone does not: binary-black-hole templates (\texttt{IMRPhenomD}~\cite{husa2016,khan2016}) match the truncated near-zone bursts in aLIGO design noise~\cite{aligo2010} with fitting factors of $\clmDetFFWindowLo$--$\clmDetFFWindowHi$, as they match the vacuum control. The vacuum control climbs the point-mass track to a peak at $fM=\clmGwTwinPeakFM$ and turns toward the equal-mass Kerr mode at $fM=\clmDetKerrFM$~\cite{berti2006,berti2009}, so its frequency rises monotonically and then stays fixed. No drainhole run follows it. The head-on's frequency falls from contact onto the $\ell=2$ quasinormal mode of a Schwarzschild black hole of its final MOTS mass, which it reaches to $\clmHeadonQnmDev\%$ at the track's minimum before a late upturn. The remnant does not spin, and the shrinking of its horizon would raise that mode, not lower it. The fly-by's frequency falls as the pair separates, and the merger's rises after contact. Wherever a remnant's ringdown is resolved (the isolated throat's, Sec.~\ref{sec:single:gw}; the head-on's and its scalar dipole, Sec.~\ref{sec:gw:censorship}), it is the ordinary quasinormal spectrum of the black hole left behind, as in any catalogue event~\cite{gwtc3}. The wormhole's imprint lies in the burst before the ringdown.

\subsection{Energy per event}
\label{sec:ligo:burst}

Table~\ref{tab:energies} lists the detector-independent result, the gravitational-wave energy per event. The background's energy density per logarithmic frequency interval in units of the critical density, $\Omega_{\rm GW}$, scales as event density times energy per event. These energies therefore turn any bound on $\Omega_{\rm GW}$~\cite{lvk2021,pagano2016}, or the rate-density limits of the LIGO--Virgo--KAGRA (LVK) all-sky burst search of the third observing run (O3)~\cite{lvk2021burst}, into a bound on the abundance of primordial drainhole encounters, if the phantom scalar exists at that epoch. The scalar radiation adds comparable energy of opposite sign (Sec.~\ref{sec:gw:scalar}).

\begin{table}[t]
\caption{Gravitational-wave energy per event in units of the source's ADM mass $M$, in the dominant multipole about the source's axis, read as in Fig.~\ref{fig:gw_ligo}(d); last column, the change from the next sphere in.}
\label{tab:energies}
\begin{ruledtabular}
\begin{tabular}{lcc}
source & $E_{\rm GW}/M$ & change (\%) \\
\hline
fly-by, $p=\clmSpiralMomentumFlyby$ & $\clmGwEnergyFlyby$ & $\clmGwSpreadFlyby$ \\
$d=6$ merger & $\clmGwEnergySpiral$ & $\clmDetESpiralSpread$ \\
head-on, $d=\clmHeadonSeparation$ & $\clmDetEHeadon$ & $\clmDetEHeadonSpread$ \\
collapsing throat & $\clmGwEnergyThroat$ & $\clmDetEThroatSpread$ \\
vacuum control & $\clmGwEnergyTwin$ & $\clmDetETwinSpread$ \\
\end{tabular}
\end{ruledtabular}
\end{table}

\section{Discussion}
\label{sec:discussion}

Every end state listed in Sec.~\ref{sec:intro} occurs, the fast plunge included, and each is selected by a measured rule. The relative scalar sign separates scattering from merger (Sec.~\ref{sec:pair}), the sign of the perturbation separates collapse from inflation (Sec.~\ref{sec:single}), and the orbital momentum separates plunge from fly-by at $\clmSpiralBoundaryLow$--$\clmSpiralBoundaryHigh\%$ of circular (Sec.~\ref{sec:spiral}). For detection, three results matter: the binaries do not inspiral (Sec.~\ref{sec:discussion:noinspiral}), the wormhole phase is a burst with no chirp ahead of it whose frequency falls or drifts until the remnant's ordinary ringdown takes over (Sec.~\ref{sec:ligo:ringdown}), and the scalar counterpart decays once a horizon forms (Sec.~\ref{sec:gw:censorship}). A wormhole signature is therefore a drifting-frequency burst with a scalar counterpart. A ringdown without a scalar counterpart comes from the black hole that remains.

\subsection{No wormhole inspiral}
\label{sec:discussion:noinspiral}

We find that a drainhole binary does not live long enough to radiate an inspiral. An inspiral needs many orbits, but each throat lives only $t\approx\tau\ln(1/\varepsilon)$, with $\tau=\clmTauLevelFourRounded$--$\clmTauRounded$ in isolation. Under the combined pull of Eq.~\eqref{eq:force}, softened as $(d+\delta)^{-2}$ at the separations evolved (Sec.~\ref{sec:pair:force}), a circular orbit at $d=12$ has period $P_{\rm orb}\approx\clmDetOrbitPeriodLo$--$\clmDetOrbitPeriodHi$. The companion acts on each throat like an inward perturbation of $\clmRaceKickTwelve\%$ (Sec.~\ref{sec:spiral:race}), which leaves a lifetime of $\clmDetCompanionLifeLo$--$\clmDetCompanionLifeHi$ units, under half a period. Even $\varepsilon=\clmDetSeedExampleC$ would give only $\clmDetLifePeriodsLo$--$\clmDetLifePeriodsHi$ periods, and each further factor of ten in $\varepsilon$ adds only $\tau\ln10\approx\clmDetDecadeUnitsLo$ units. The period grows with separation as $d^{3/2}$ and the lifetime only as $\ln(1/\varepsilon)$, and no binary is free of its companion's perturbation.

Radiation reaction does not help either. The leading radiative moment, the scalar dipole, carries negative energy comparable to the gravitational radiation (Sec.~\ref{sec:gw:scalar}), which, if the ADM balance holds, pumps the orbit rather than draining it. Pumping acts over many orbits, whereas each encounter evolved here is settled within one passage by the conservative pull and contact (Sec.~\ref{sec:spiral:capture}). Over the fly-by the scalar radiation carries only $\clmGwScalarRatioFull$ of the gravitational energy (Sec.~\ref{sec:gw:scalar}), so a close pass loses energy on net. Pumping matters for the slow, wide orbits an inspiral needs, where the dipole dominates as the lower multipole, and no throat survives to reach them.

The signal specific to wormholes is therefore a short transient. Within a few tens of $M$ each mouth collapses or inflates, so the plunge, burst and collapse to a black hole of Secs.~\ref{sec:headon}--\ref{sec:gw} are all this source emits while its throats are open. Any long chirp belongs to the objects the throats become. The argument holds while the throat's unstable mode exists, as for the minimally coupled phantom scalar without spin of the Gonz\'alez--Guzm\'an--Sarbach analysis. Rotation can remove that mode, in five dimensions above a critical angular momentum~\cite{dzhunushaliev2013} and in four as perturbative analyses indicate~\cite{azad2024plb,azad2024prd,khoo2024}; spinning Ellis wormholes exist up to an extremal limit~\cite{kleihaus_kunz2014}. Other supports are radially stable~\cite{poisson_visser1995,kanti2011}, so a spinning or differently supported pair that outlives an inspiral is not excluded here. The same lifetimes rule out a static phantom-supported throat as a long-lived black-hole mimicker of the kind ringdown and echo tests target~\cite{cardoso2016,bueno2018}.

\begin{figure*}[t]
\centering
\includegraphics[width=\textwidth]{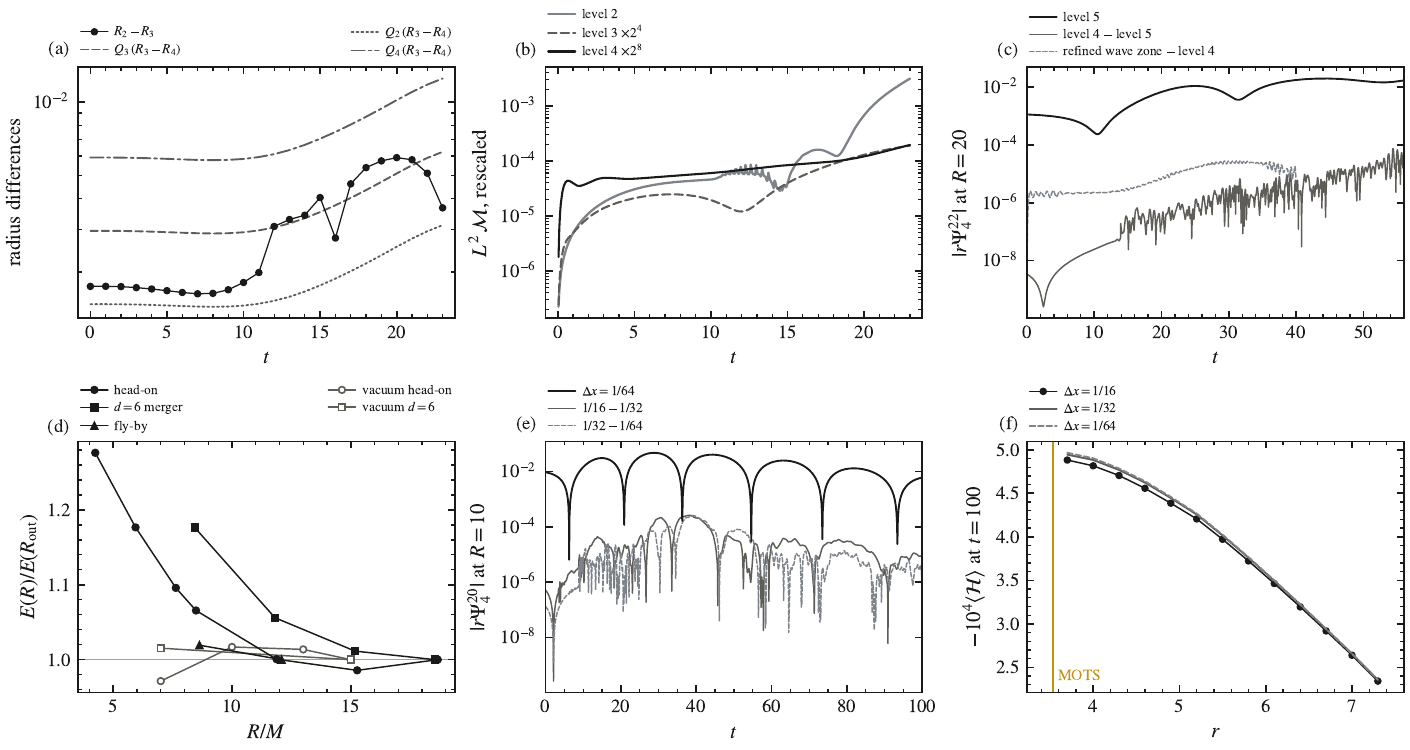}
\caption{(a)~The static throat from exact data at levels 2, 3 and 4: the difference $R_2-R_3$ of its minimal areal radius between levels 2 and 3 (ink), and $R_3-R_4$ multiplied by $Q_n=2^n$ for $n=2$, 3, 4 (grey), against time; the curve that overlays $R_2-R_3$ gives the order, and the dip at $t=16$ is one level-2 output time. (b)~The $L^2$ norm of the momentum constraint for the same runs, multiplied by $2^4$ for each level above level 2, the factor for fourth-order convergence. (c)~$|r\Psi_4^{22}|$ at $R=\clmExtractRadiusTwenty$ for the $d=12$, $p=\clmSpiralMomentumProd$ binary from Bowen--York initial data (not the boosted data used elsewhere) at throat level 5, its difference from level 4, and the change from refining the wave zone to half the base cell. (d)~Radiated energy on each sphere over that on the outermost sphere, against $R/M$, for the head-on run, the $d=6$ merger and the fly-by (ink) and two vacuum controls (grey). (e)~$|r\Psi_4^{20}|$ at $R=\clmExtractRadiusTen$ for the head-on runs in the larger domain of Fig.~\ref{fig:gw_gallery}(b), at finest $\Delta x=1/\clmConvFzDxLo$, $1/\clmConvFzDxMid$ and $1/\clmConvFzDxHi$, and each neighbouring pair's difference (at most $\clmConvFzPsiDiff\%$, then $\clmConvFzPsiDiffFine\%$, of the peak); all three form the common MOTS at $t=\clmConvFzMotsTime$. (f)~Shell averages of $\mathcal{H}_{\rm ADM}$ against $r$ outside the common MOTS (gold rule, its largest coordinate extent) for the same three runs at $t=100$, on the shared level-4 grid. The static throat converges at order $\clmConvOrderEvolLo$--$\clmConvOrderEvolHi$ and its momentum constraint between third and fourth order, the head-on energy at order $\clmConvFzOrder$ and $\mathcal{H}_{\rm ADM}$ at order $\clmConvFzHamOrder$, so the interior's constraint violation does not leak past the horizon.}
\label{fig:convergence}
\end{figure*}

\subsection{Topological censorship}
\label{sec:discussion:topology}

Topological censorship forbids an observer at infinity from probing the topology of a globally hyperbolic, asymptotically flat spacetime that satisfies the null energy condition~\cite{friedman1993}. A traversable wormhole evades it only by violating that condition, and the two branches end the evasion in opposite ways. On the collapse branch, alone or in a merger, the throats end inside a MOTS while the phantom field outside is swallowed or radiated away (Secs.~\ref{sec:single:collapse}, \ref{sec:headon:wall} and~\ref{sec:gw:censorship}). The wormholes are not removed (no classical evolution changes the number of asymptotic ends~\cite{geroch1967}) but end hidden behind a horizon, as the theorem would have them. On the inflation branch no trapped surface forms and the throat stays open (Sec.~\ref{sec:single:fate}). The evasion persists, and a distant observer meets an expanding anti-trapped region with no horizon in front of it up to the end of the run.

\subsection{Scope}
\label{sec:discussion:open}

Our results rest on one matter model and one family of data: a single minimally coupled phantom scalar, equal masses, one throat compactness ($m/a=1/2$), no spin, and constraint-solved, matched initial data whose momentum is an exact boost (Secs.~\ref{sec:model:solve} and~\ref{sec:model:boost}). Horizons here are MOTSs on the evolved slices, which for matter violating the null energy condition need not lie inside event horizons (Sec.~\ref{sec:setup:mots}). Within this family the instability rules out survival from an early formation epoch to late times. A long-lived relic needs spin or a radially stable support (Sec.~\ref{sec:discussion:noinspiral}), both outside this family. Nonlinear evolutions of rotating throats find that rotation slows the radial instability and, above a threshold spin, show no evidence of it over the evolved times~\cite{shirokov2026c}.

\section{Reproducibility and cost}
\label{sec:repro}
\label{sec:cost}

Every figure is regenerated by a script from the archived diagnostic time series of the evolutions in Table~\ref{tab:matrix}; the archive and the code are listed under Declarations. The evolutions are bit-for-bit reproducible on the same hardware: a single-throat run repeated with checkpoint output gives byte-identical diagnostics over $\clmControlByteRows$ output times.

Every evolution ran on one NVIDIA H100 \clmDetMemDevice\,GB device, with \texttt{GRTeclyn} built CUDA-native against AMReX~\clmDetAmrexVersion. The $\clmDetRunsTotal$ evolutions total $\clmDetGpuHours$ GPU-hours, each run's cost recomputed from the speed and time span in its archived logs.

\begin{acknowledgments}
The authors acknowledge the GRTL Collaboration for \texttt{GRChombo}, \texttt{GRTresna},
and \texttt{GRTeclyn}.
\end{acknowledgments}

\section*{Declarations}
\textit{Funding.} This work was supported by Gravity Frontiers (\url{https://www.gravityfrontiers.org/en}), which funded the underlying research and the development of the numerical-simulation software.

\textit{Competing interests.} The authors declare no competing interests.

\textit{Data availability.} The parameter files, diagnostic time series and figure scripts of every simulation are archived at \url{https://doi.org/10.5281/zenodo.23257999}.

\textit{Code availability.} The evolution and analysis code is the \texttt{GRTeclyn}~\cite{grteclyn} fork at \url{https://github.com/Nikchik-coder/GRTeclyn}, tag \texttt{v1.0.1-wormhole-merger}, archived with the data. It contains the drainhole initial data, constraint solve and boost, the perturbations of Sec.~\ref{sec:model:seeds}, the MOTS finder, the extraction and figure scripts, and a matched-filter search of public strain~\cite{gwosc,pycbc}.

\FloatBarrier

\appendix

\section{Convergence and constraint monitoring}
\label{app:health}
\packfloatpages

\begin{figure*}[t]
\centering
\includegraphics[width=0.80\textwidth]{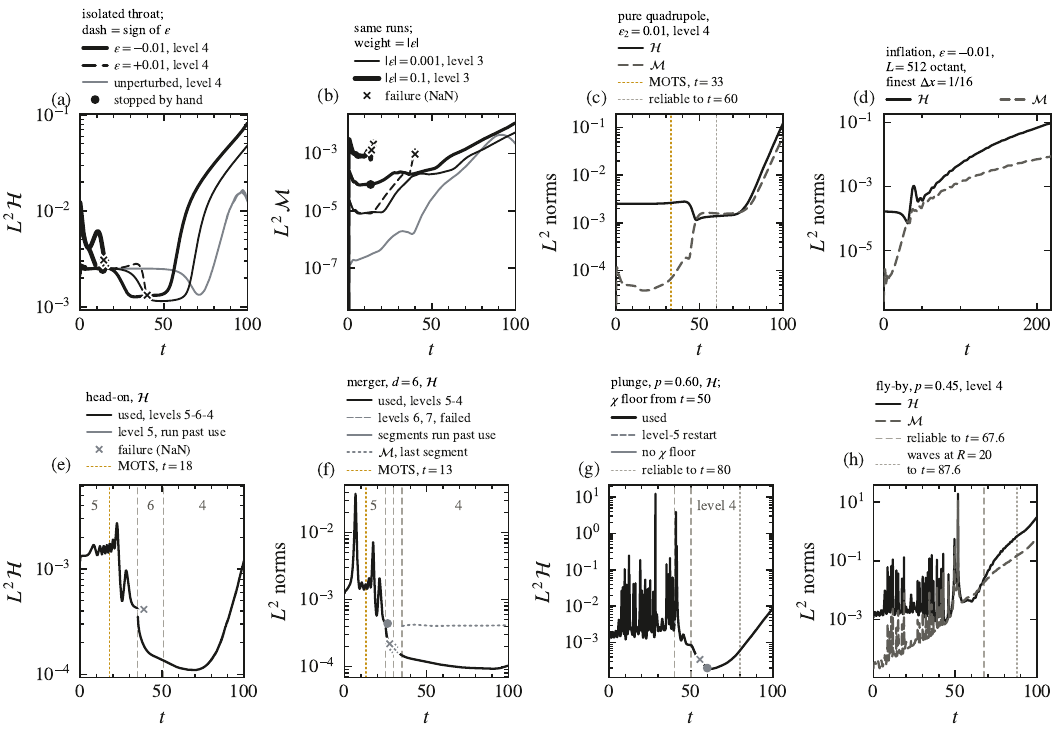}
\caption{$L^2$ norms of the Hamiltonian ($\mathcal{H}$, ink) and momentum ($\mathcal{M}$, grey dashed) constraints against time for every run drawn in the main-text figures, each at its highest refinement level. The norms are base-grid averages and compare only within one domain (Sec.~\ref{sec:setup:diagnostics}). (a),(b)~The single throat of Fig.~\ref{fig:single_throat}: level 4 for the unperturbed throat and $\varepsilon=\pm10^{-2}$ ($+10^{-2}$ stopped, dot), level 3 for $\pm10^{-3}$ and $\pm0.1$ (dash style gives the sign, line weight the amplitude; grey, unperturbed; crosses, failures). At $t=0$ the norms carry the discretisation error of the exact data plus the perturbation's own constraint violation, $\clmSeedDefectBox\,|\varepsilon|$ (Sec.~\ref{sec:model:seeds}). The runs with $|\varepsilon|\le10^{-2}$ then follow the unperturbed norms until the end states part; $\varepsilon=\pm0.1$ starts higher and fails by overflow at the compactified origin. (c)~The pure-quadrupole collapse of Fig.~\ref{fig:single_collapse} (gold, its MOTS from $t=\clmPureQuadMotsTime$; dotted, reliable to $t=\clmPureQuadTrust$); the late rise comes after that time. (d)~The inflating throat of Fig.~\ref{fig:single_inflation} to $t=\clmFourWall$. (e)~The head-on run of Fig.~\ref{fig:headon_collapse} (rules at the restart times, levels per restart segment). Its level-5 segment fails (a NaN inside the horizon) at $t=\clmHeadonCsmNanTime$ (cross), with its norm below its value at horizon formation (gold rule, $t=\clmHeadonCsmMotsTime$). (f)~The $d=6$ merger of Fig.~\ref{fig:spiral_collapse}: the restart segments used in ink, each failed restart in grey up to its failure (levels 6 and 7 dashed; times in Fig.~\ref{fig:spiral_collapse}(d)). (g)~The $p=\clmEgwPSixty$ plunge at level 4: the run without a $\chi$ floor (grey) fails by a NaN in the merged core at $t=\clmPSixtyExtDeath$ with its norm at its floor, the restart with a $\chi$ floor reaches $t=100$, and a level-5 restart (grey dashed) passes the merger (dotted, reliable to $t=\clmPSixtyTrust$). (h)~The $p=\clmSpiralMomentumFlyby$ fly-by: the norms stay at their floor through the pass and grow as the mouths inflate (rule, reliable to $t=\clmPFortyFiveTrust$). The short peaks in (g), (h) and the $p=\clmSpiralMomentumTwentyFive$ run come from the base-grid norm sampling the moving throats' $\chi$ minima: on the $p=\clmSpiralMomentumTwentyFive$ test pair, four cells next to the minima carry $\clmNormPitShare\%$ of a peak, and neither $\max|K|$ nor the waves show it. At every failure the Hamiltonian norm sits at or below its starting value, so each failure is the interior NaN of Sec.~\ref{sec:spiral:wall}, not a constraint blow-up.}
\label{fig:constraints}
\end{figure*}

Figure~\ref{fig:convergence} tests convergence. For grid spacings $h_1>h_2>h_3$, a quantity $f$ that converges at order $n$ has $(f_1-f_2)/(f_2-f_3)=Q_n\equiv(h_1^{\,n}-h_2^{\,n})/(h_2^{\,n}-h_3^{\,n})$, and $Q_n=2^n$ when each level halves the cell. The static throat, the one configuration with an exact answer, is evolved at levels 2, 3 and 4. Its minimal areal radius converges at order $\clmConvOrderDataLo$--$\clmConvOrderDataHi$ while the error of reading the exact data dominates and at order $\clmConvOrderEvolLo$--$\clmConvOrderEvolHi$ once the evolution's own error dominates (Fig.~\ref{fig:convergence}(a)), between the fourth order of the stencils and the second of the time interpolation at refinement boundaries. Rescaling a constraint norm by $2^n$ between two resolutions is valid because its exact value is zero. The momentum constraint vanishes identically for the static throat, so its norm measures only the error the evolution adds. Rescaled for fourth order, the norms at the three levels agree to within a small factor, a convergence between third and fourth order (Fig.~\ref{fig:convergence}(b)). The Hamiltonian-constraint norm, a base-grid average, is the same at all three levels, since the base grid's own error in the static data sets it. The Bowen--York $d=12$ binary of Fig.~\ref{fig:convergence}(c) at throat levels 4 and 5 agrees in $r\Psi_4$ to $\clmConvPsiLevelDiff\%$ of its burst peak, and refining its wave zone moves it by $\clmConvPsiWaveZoneDiff\%$.

Figure~\ref{fig:constraints} shows the constraint norms of every run drawn in the main-text figures. The norms are volume averages on a base grid that does not resolve the throats (Sec.~\ref{sec:setup:diagnostics}), so the throats set their absolute scale, and the norms serve as monitors, not as wave-zone error bars. At every failure the norm sits at or below its earlier value. The head-on runs in the larger domain ($L=\clmFarzoneBoxL$) of Fig.~\ref{fig:convergence}(e),(f) test whether the interior affects the waves. These three runs differ only in the throats' finest cell (all with $\chi\ge10^{-4}$ and $\sigma_{\rm KO}=0.3$, unlike the main head-on run; Table~\ref{tab:mods}). They agree pairwise to sub-percent level in the exterior waveform, the horizon track, the radiated energy and the Hamiltonian constraint outside the MOTS, so the interior's resolution does not affect the exterior.

\begin{figure*}[t]
\centering
\includegraphics[width=0.62\textwidth]{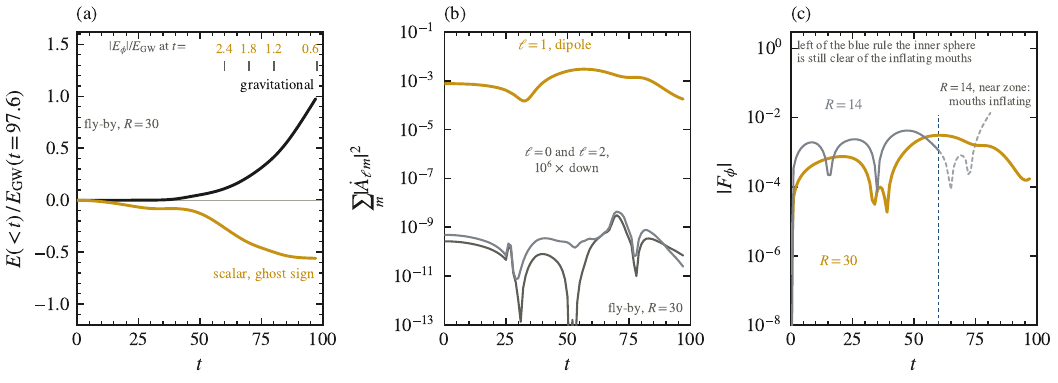}\\[4pt]
\includegraphics[width=0.62\textwidth]{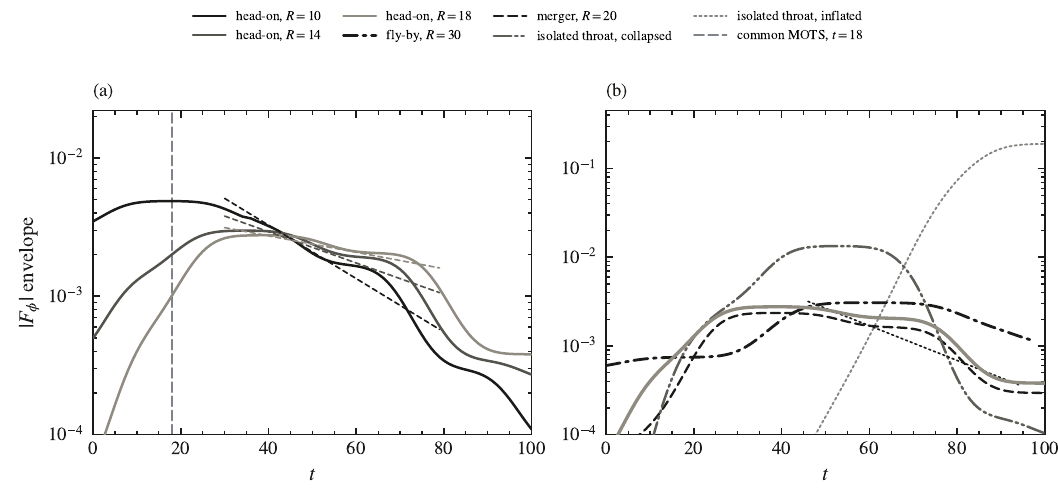}
\caption{Top: energy, multipoles and flux of the scalar field on the $\Psi_4$ extraction spheres, for the fly-by.
(a)~Energy through $R=\clmGwScalarRadiusOuter$ against time, in units of the gravitational energy that has crossed it by $t=\clmGwScalarGateOuter$ (reliable to $t-R=\clmFlybyTrustRetarded$): gravitational (ink) and scalar with its physical sign (gold, Sec.~\ref{sec:gw:scalar}). The printed ratio $|E_\phi|/E_{\rm GW}$ falls to $\clmGwScalarRatioFull$ over the whole record.
(b)~Multipole content: $\ell=1$ carries all but $10^{-\clmGwScalarDipoleShare}$ of the power; $\ell=0$ and $\ell=2$, forbidden by symmetry, measure numerical symmetry breaking.
(c)~$|F_\phi|$ at both spheres, each to $t-R=\clmFlybyTrustRetarded$. After the blue rule ($t=\clmGwScalarCut$) the inner sphere reads the inflating mouths' near field, which has no counterpart at $R=\clmGwScalarRadiusOuter$ one light-crossing time later, so that stretch is dashed. The scalar energy leaves as a dipole pulse that arrives before the gravitational burst.
Bottom: envelope of the scalar flux $|F_\phi|$ against time, for runs with and without a horizon: a running maximum over $\clmGwEnvWindow$ units, smoothed in log amplitude, since the raw flux changes sign every ${\sim}\clmGwFluxSignStretch$ units.
(a)~The head-on run (all restart segments joined) at $R=\clmGwHeadonRadiusA/\clmGwHeadonRadiusB/\clmGwHeadonRadiusC$. The common MOTS forms at $t=\clmHeadonCsmMotsTime$ (dotted). The remaining scalar charge then leaves as one $\ell=1$ pulse, peaking later at each sphere, and decays (dashed, exponential fits). Before the rule the flux, from the open mouths' static scalar charge, is canonically ingoing; only the outgoing part is integrated.
(b)~One sphere per end state (one extraction radius each). The merger, behind its horizon from $t=\clmMergerMotsTime$, decays like the head-on (dotted fit). The fly-by, without a horizon, shows its complete single pulse (reliable to $t-R=\clmFlybyTrustRetarded$). Mid-grey: the single throat twice, the $\varepsilon=\clmGwKickedEps$ level-\clmGwKickedLevel{} collapse (solid, MOTS from $t=\clmGwKickedMotsTime$) and the unperturbed level-\clmGwSingleLevel{} throat (dotted), which inflates without a horizon and grows in the monopole. A horizon switches the scalar flux off exponentially; without one, the flux ends only with the encounter, or grows while a throat inflates.}
\label{fig:scalar_channel}\label{fig:scalar_censorship}
\end{figure*}

\section{Scalar radiation}
\label{app:supp}

Figure~\ref{fig:scalar_channel} shows the radiated scalar energy, its multipoles and the time range over which it is reliable (Sec.~\ref{sec:gw:scalar}), and how a horizon shuts off the scalar flux (Sec.~\ref{sec:gw:censorship}).

\FloatBarrier

\end{document}